\documentclass[12pt]{iopart}
\usepackage{graphics}
\usepackage{epsfig}
\usepackage{epstopdf}
\usepackage{cite} 
\newcommand{\au}{~a.u.}

\newcommand{\eqn}[1]{(\ref{#1})}
\newcommand{\fig}[1]{figure \ref{#1}}
\newcommand{\tab}[1]{table~\ref{#1}}
\newcommand{\secref}[1]{setion \ref{#1}}

 \usepackage{hyperref}
 \hypersetup{
    unicode=false,          
    pdftoolbar=true,        
    pdfmenubar=true,        
    pdffitwindow=false,     
    pdfstartview={FitH},    
    pdftitle={Positron scattering and annihilation in hydrogen-like ions},    
    pdfauthor={D. G. Green},     
    pdfsubject={Positron_hlikeions},   
    pdfcreator={D. G. Green},   
    pdfproducer={Producer}, 
    pdfkeywords={keywords}, 
    pdfnewwindow=true,      
    colorlinks=true,       
    linkcolor=blue,          
    citecolor=blue,        
    filecolor=magenta,      
    urlcolor=blue           
}

\usepackage{iopams}  

\begin{document}
\title[Effect of positron-atom interactions on the annihilation gamma spectra of molecules]{Effect of positron-atom interactions on the annihilation gamma spectra of molecules}

\author{D~G~Green$^1$, S~Saha$^2$, F~Wang$^2$, G~F~Gribakin$^1$ and C~M~Surko$^3$}
\address{$^1$ Department of Applied Mathematics and Theoretical Physics, Queen's University Belfast, Belfast, BT71NN, Northern Ireland, United Kingdom }
\address{$^2$ Faculty of Life and Social Sciences, Swinburne University of Technology, Hawthorn, Victoria 3122, Australia}
\address{$^3$ Physics Department, University of California, San Diego, La Jolla, California 92093-0319, USA}
\ead{dermot.green@balliol.oxon.org}

\begin{abstract}
Calculations of $\gamma$-spectra for positron annihilation on a selection of molecules, including methane and its fluoro-substitutes, ethane, propane, butane and benzene are presented. 
The annihilation $\gamma$-spectra characterise the momentum distribution of the electron-positron pair at the instant of annihilation. The contribution to the $\gamma$-spectra from individual molecular orbitals is obtained from electron momentum densities calculated using modern computational quantum chemistry density functional theory tools. The calculation, in its simplest form, effectively treats the low-energy (thermalised, room-temperature) positron as a plane wave and gives annihilation $\gamma$-spectra that are  about 40\% broader than experiment, although
the main chemical trends are reproduced. 
We show that this effective ``narrowing'' of the experimental spectra is due to the action of the molecular potential on the positron, chiefly, due to the positron repulsion from the nuclei.  It leads
to a suppression of the contribution of small positron-nuclear separations where the electron momentum is large. 
To investigate the effect of the nuclear repulsion, as well as that of short-range electron-positron and positron-molecule correlations, a linear combination of atomic orbital description of the molecular orbitals is employed. 
It facilitates the incorporation of correction factors which can be calculated from atomic many-body theory and account for the repulsion and correlations.
Their inclusion in the calculation gives $\gamma$-spectrum linewidths that are in much better agreement with experiment.
Furthermore, it is shown that the effective distortion of the electron momentum density, when it is observed through positron annihilation $\gamma$-spectra, can be approximated by a relatively simple scaling factor. 
\end{abstract}

\submitto{\NJP}
\maketitle

\section{Introduction}\label{intro}
In the dominant positron annihilation process in matter, a positron annihilates with an electron to produce two gamma ($\gamma$) photons with a total energy of $2mc^2$ and a total momentum ${\bf P}$~\cite{Dirac_annihilation,qed}. 
Nonzero values of {\bf P} lead to Doppler shifts of the photon energies from $mc^2\approx511$\,keV.  
In the case of bound electrons, this Doppler broadening is mostly due to the electron momenta~\cite{PhysRevA.55.3586, PhysRevLett.79.39,0953-4075-29-12-004} ---
the spectra of $\gamma$-ray energies (the `$\gamma$-spectra') therefore contain information about the electron momentum distribution in the bound-state orbitals and can be used to characterize the orbitals involved~\cite{0953-4075-39-7-008,DGG_hlike,DGG_innershells, wangnobles}.

In light of this fact, positron annihilation spectroscopy is a valuable technique in the study of surfaces, interfaces and defects in condensed matter~\cite{RevModPhys.60.701, RevModPhys.66.841}, of electronic properties of quantum dots~\cite{PhysRevB.65.245310,PhysRevB.66.041305,PhysRevB.68.165326,nmat1550, eijt:091908, meng:093510}, nanoparticles~\cite{PhysRevB.79.201405, PhysRevB.73.014111} and nanocrystals~\cite{positronnanocrystals}, and in the study and exploitation of positron annihilation on core electrons~\cite{PhysRevLett.77.2097,PhysRevLett.82.3819, PhysRevB.51.4176,PhysRevB.73.014114, paescopper,PhysRevLett.79.39, DGG_innershells}.

In all of these areas, interpretation of the measured annihilation spectra relies heavily on theoretical input. 
However, the problem of calculating $\gamma$-spectra for positron annihilation on molecules has received little attention to date, with previous positron-molecule theoretical studies focussed mainly on understanding the anomalously large annihilation rate parameters $Z_{\rm eff}$~\cite{RevModPhys.82.2557}.
This is in contrast with condensed matter systems, where a number of theoretical approaches have been developed~\cite{RevModPhys.60.701, RevModPhys.66.841}, and atomic systems, for which detailed theoretical understanding has been provided by, e.g., elaborate variational~\cite{0953-4075-29-12-004} and many-body theory~\cite{0953-4075-39-7-008,DGG_hlike, DGG_innershells} calculations. 
The aim of this paper is to present calculations of $\gamma$-spectra for positron annihilation on a range of molecules including hydrogen, methane and its fluoro-substitutes, ethane, propane, butane and benzene.

The first measurement of the Doppler-broadened $\gamma$-spectrum for positron annihilation on molecules was made in 1986 by Brown and Leventhal in low-density hydrogen gas~\cite{PhysRevLett.57.1651}. 
In the early 1990's the confinement of thermalized positrons in a Penning trap allowed Tang \emph{et al.}~\cite{PhysRevLett.68.3793} to measure $\gamma$-spectra for a range of molecules, including hydrocarbons and perfluorocarbons. 
This work was built upon by Iwata \emph{et al.}~\cite{PhysRevA.55.3586} who measured $\gamma$-spectra for positron annihilation on a large range of atoms and molecules in the gas phase. 
Despite this extensive set of experimental results, there is a paucity of theoretical calculations of the annihilation $\gamma$-spectra in molecules. 
Theoretical predictions of the $\gamma$-spectra do exist for the molecular ion H$_2^+$, for which (numerical) Coulomb-Born and configuration-interaction Kohn variational calculations have been performed~\cite{0953-4075-30-6-025}, and for the small molecules H$_2$ and N$_2$~\cite{darewych,ghosh}. 
However, the application of these sophisticated methods to larger molecules is considerably more difficult.
More than thirty years ago Chuang and Hogg~\cite{chuang} analysed the annihilation $\gamma$-spectra for hexane and decane.
However, their calculation for the alkane molecules relied on analytic wave functions of the carbon orbitals calculated in the 1950's, which are relatively crude by modern standards. Furthermore, their calculations disregarded the effect of the positron wavefunction. New calculations using modern quantum chemistry methods, similar to those reported here, are underway \cite{fluorobenzenes}.

Although the $\gamma$-spectra are determined primarily by the electronic structure, they are also sensitive to the positron interaction with the target. 
Compared to positron annihilation on atoms, the positron-molecule problem is significantly more complex owing to, e.g., the non-spherical potential experienced by the positron. 
In this work we use two methods to calculate $\gamma$-spectra for positron annihilation on molecules.
The first method is based on electron momentum densities calculated from modern computational chemistry tools, such as density functional theory (DFT). 
This approach was recently tested by a number of the authors in a study of positron annihilation on noble gases~\cite{wangnobles} and applied to
fluorobenzenes \cite{fluorobenzenes}.
In its simplest form, this method treats the positron as a plane wave and neglects its influence on the $\gamma$-spectrum linewidth. 
To overcome this deficiency, and to investigate the effect of the atomic potentials and electron-positron correlations, we perform additional calculations that rely on treating the molecular orbitals as linear combinations of atomic orbitals.
This yields predicted spectra that are in good agreement with the computational chemistry calculations and furthermore, it gives a framework for the inclusion of the effect of the atomic potentials and correlations on the $\gamma$-spectrum linewidth (see Ref.~\cite{GSW11} for preliminary results).
 
The structure of the paper is as follows. 
Section~\ref{sec:theory} briefly recaps the general theory of positron annihilation $\gamma$-spectra.
In section~\ref{sec:dft} the results of the DFT calculations are presented for molecular hydrogen, methane and its fluorosubstitutes, ethane, propane, butane and benzene. 
This DFT calculation effectively ignores the effect of the positron-molecule interaction on the $\gamma$-spectra, and gives $\gamma$-spectra that are systematically broader than experiment.
In section~\ref{sec:LCAO} an investigation of the systematic broadening is performed. 
There, we derive the form of the annihilation $\gamma$-spectra for molecules using a multicentre linear combination of atomic orbital approach. 
We also show how, in this scheme, the effect of the nuclear repulsion and electron-positron correlations can be accounted for through momentum-dependent correction factors obtained from positron-atom calculations, e.g., using many-body theory.
This procedure is applied to H$_2$, CH$_4$ and CF$_4$ in section~\ref{sec:results}.
In light of our results, we discuss how the effective distortion of the electron momentum density, when it is observed through positron annihilation $\gamma$-spectra, can be approximated by a relatively simple energy scaling factor. 
We conclude with a summary in section~\ref{sec:conclusions}.

\section{Theory: gamma-spectra for positron annihilation on molecules}\label{sec:theory}
\subsection{Basic equations}
Low-energy positrons annihilate in matter predominately via two-photon production, a process in which the total spin of the electron-positron pair must be zero~\cite{Akhiezer,qed}.
In the centre-of-mass frame of the pair, where the total momentum of the two photons ${\bf P}={\bf p}_{\gamma_1}+{\bf p}_{\gamma_2}$ is zero, the two photons propagate in opposite directions and have equal energies $E_{\gamma}=p_{\gamma}c=mc^2+\frac{1}{2}(E_i-E_f)\approx mc^2=511~{\rm keV}$, where $E_i$ and $E_f$ are the energies of the initial and final states (excluding rest mass),  respectively. When ${\bf P}$ is nonzero, however, the two photons no longer propagate in exactly opposite directions and their energies are Doppler shifted.
For example, for the first photon $E_{\gamma_1}=E_{\gamma}+m{\bf c}\cdot{\bf V}$, where ${\bf V}={\bf P}/2m$ is the centre of mass velocity of the electron-positron pair and ${\bf c}=c\,\hat {\bf c}$, where $\hat{\bf c}$ is the unit vector in the direction of the photon.
Assuming that $V\ll c$, and $p_{\gamma_1}=E_{\gamma_1}/c\approx mc$, the shift of the photon energy from the centre of the line, $\epsilon\equiv E_{\gamma_1}-E_{\gamma}$, is
\begin{eqnarray}
\epsilon=m{\bf c}\cdot{\bf V}=\frac{1}{2}{\bf P}\cdot{\bf c}.
\end{eqnarray}
When a low-energy positron annihilates with a bound electron, the characteristic Doppler shifts are determined by the electron energy $\varepsilon_n$, $\epsilon \sim\sqrt{|\varepsilon_n|mc^2}\gg|\varepsilon_n|$. 
Hence, the shift of the line centre $E_\gamma $ from $mc^2=511$~keV can usually be neglected.
The annihilation $\gamma$-spectrum (Doppler spectrum) can then be written in a form similar to a Compton profile, as the integral of the annihilation probability density
\begin{eqnarray}
w_{fi}(\epsilon, \hat{\bf c})=\int \left|A_{fi}({\bf P})\right|^2 \delta\left(\epsilon-\frac{1}{2}\bf {P}\cdot{\bf c}\right) \frac{d^3P}{(2\pi)^3},
\end{eqnarray}
where $A_{fi}({\bf P})$ is the amplitude for annihilation into two photons. 
It is the fundamental quantity to be evaluated. 
A many-body theory description of it is discussed below.

For gaseous systems, the measured spectrum represents an average of $w_{fi}(\epsilon, \hat{\bf c}) $ over the direction of emission of the annihilation photons~\cite{0953-4075-39-7-008,DGG_innershells}
\begin{eqnarray}\label{eqn:spectra}
w_{fi}(\epsilon)
=\frac{1}{c}\int _{2|\epsilon|/c}^{\infty}\rho_{fi}^{a}(P) \,4\pi{PdP},
\end{eqnarray}
where 
\begin{eqnarray}\label{eq:rhoaP}
\rho_{fi}^{a}(P)\equiv\int \frac{\left|A_{fi}({\bf P})\right|^2}{(2\pi)^3}\frac{d\Omega_{\bf P}}{4\pi},
\end{eqnarray}
is the spherically averaged annihilation momentum density. 
If the initial state corresponds to a positron with momentum ${\bf k}$ incident on the
ground-state target, and the possible final states describe the target with an electron missing
in orbital $n$, the total spectrum is found by summing over all final states as
\begin{eqnarray}\label{eqn:spectra_sumf}
w_{{\bf k}}(\epsilon)=\sum_nw_{n{\bf k}}(\epsilon)
=\frac{1}{c}\int _{2|\epsilon|/c}^{\infty}\rho_{{\bf k}}^{a}(P) \,4\pi{PdP},
\end{eqnarray}
where $\rho_{{\bf k}}^{a}(P)=\sum_n\rho_{n{\bf k}}^{a}(P)$ is the total annihilation momentum density that is equal to the sum of the individual annihilation momentum densities of the occupied molecular orbitals $n$.
The probability of positron annihilation with core electrons is relatively small (owing to the positron repulsion from the nuclei). Hence, the annihilation spectra are dominated by the contribution of valence orbitals, except at high Doppler shifts where a distinct contribution from the core can be seen \cite{PhysRevLett.79.39,DGG_innershells}.

%
\subsection{The annihilation amplitude}
\subsubsection{General form}
The annihilation amplitude that enters \eqn{eqn:spectra} is defined as
\begin{eqnarray}
A_{n\bf k}({\bf P})=\left\langle \Psi^{N-1}_n \left|\int e^{-i{\bf P}\cdot {\bf r}}\hat \psi ({\bf r})\hat\varphi ({\bf r})d^3r \right| \Psi^{N+1}_{\bf k}\right\rangle,
\end{eqnarray}
where $\hat \psi ({\bf r})$ and $\hat \varphi ({\bf r})$ are the electron and positron annihilation operators, respectively, $\Psi^{N+1}_{\bf k}$ is the (fully-correlated) initial state of $N$ electrons and a positron of momentum ${\bf k}$, and $\Psi^{N-1}_n$ is the (fully-correlated) final state of the target with $N-1$ electrons and a hole in electron orbital $n$.
It can be calculated using many-body theory methods (see Refs.~\cite{0953-4075-39-7-008,DGG_hlike, DGG_innershells} for details). 

\begin{figure}[ht!]
\centering
\includegraphics[width=0.75\textwidth]{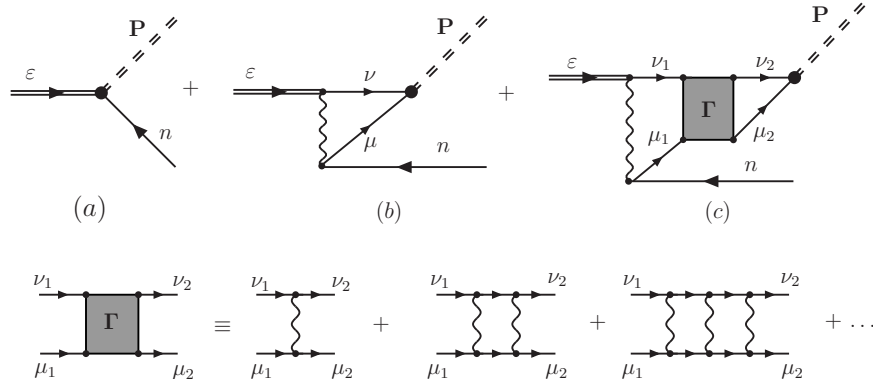}\\
\caption{Main contributions to the annihilation amplitude $A_{n{\bf k}}({\bf P})$: (a) zeroth-order (independent-particle model); (b) first-order correction; (c) $\Gamma$-block (electron-positron ladder series) correction.
All lines correspond to particles (or holes) propagating on top of the ground state of the $N$-electron system. The fully-correlated incident positron wavefunction (Dyson orbital) is represented by the double line labelled $\varepsilon$; the wavefunction of the annihilated electron (hole)  is labelled by $n$; the double-dashed lines represent the two photons of momentum
${\bf P}$; wavy lines represent Coulomb interactions; and $\nu$ and $\mu$ are excited positron and electron states, respectively.}
\label{fig:anndiagrams}
\end{figure}

Figure~\ref{fig:anndiagrams} shows the many-body diagrammatic expansion for the amplitude
(see the figure caption for an explanation of the diagrams).
More specifically, it shows the main contributions to the annihilation vertex that describes the short-range electron-positron interaction. 
Figure \ref{fig:anndiagrams}\,(a) is the zeroth-order, or  independent-particle-model (IPM) amplitude, which is given by
\begin{eqnarray}\label{eq:A0}
A_{n{\bf k}}^{(0)}({\bf P})=\int e^{-i{\bf P}\cdot{\bf r}} \psi_n({\bf r})\varphi_{\bf k}({\bf r})\,d^3r,
\end{eqnarray}
where $\varphi_{\bf k}$ is the wavefunction of the positron, and $\psi _n$ is the wavefunction of the annihilated electron (hole) in state $n$.
Figures~\ref{fig:anndiagrams}\,(b) and (c) describe the short-range electron-positron correlation corrections to the zeroth-order vertex.
Their explicit forms can be found in Refs.~\cite{0953-4075-39-7-008,DGG_hlike,DGG_innershells}.
In addition to the short-range electron-positron correlations, there are also correlations between the incident positron and the target, e.g., through a polarisation interaction. 
These interactions are described by a non-local and enegy dependent positron self-energy~\cite{PhysRevA.70.032720,0953-4075-39-7-008,DGG_hlike,DGG_innershells}. They are included
in the fully-correlated (or `dressed') incident positron wavefunction known as the
\emph{Dyson orbital}, which is represented by the double line labelled $\varepsilon$ in \fig{fig:anndiagrams}.

When evaluating the diagrams beyond zeroth order, one must sum over the complete set of intermediate electron and positron wavefunctions, including those in the continuum. 
This summation can be performed relatively easily for atomic systems, since the spherical potential allows for the wavefunctions to be separated into radial and angular parts and asymptotic convergence formulae can be used~\cite{0953-4075-35-2-311,PhysRevA.70.032720,0953-4075-39-7-008,DGG_hlike, DGG_innershells}. 
However, the lack of spherical symmetry in the molecular problem makes the development of a numerical scheme in which the higher-order diagrams can be calculated a more difficult
task.\footnote{In evaluating the diagrams for a molecular system one would have to, e.g., transform the Hartree-Fock eigenbasis to a (non-orthogonal) one that involves (at least) the atomic orbitals that constitute the molecule, centred on their respective atomic sites, as is standard in the derivation of the Roothaan equations. Alternatively, one can use a Gaussian basis.}
Nevertheless, the shape of the $\gamma$-spectra is described to a good first approximation by the zeroth-order amplitude~\cite{PhysRevLett.79.39,0953-4075-39-7-008,DGG_hlike,DGG_innershells}, the calculation of which requires only accurate positron and electron wavefunctions.
We will use this as our starting point.

To go beyond the zeroth-order approximation and incorporate the effects of the higher-order diagrams on the lineshape,  we will assume that the positron annihilates in the vicinity of one of the atoms of the molecule.
The annihilation amplitude diagrams can then be calculated using positron and electron orbitals in the field of the \emph{atomic} potentials.

\subsubsection{Plane-wave positron, independent-particle approximation}

If one assumes that the positron is described by a plane wave, then in the low-energy limit $k\ll1$\,a.u., one has $\psi_{\bf k}({\bf r})=e^{i{\bf k}\cdot {\bf r}}\approx1$, for the range of distances at which annihilation occurs.\footnote{Annihilation occurs at distances of the size of the atom or molecule involved.}
In this case the zeroth-order annihilation momentum density [Eqs.~(\ref{eq:rhoaP}) and (\ref{eq:A0})] reduces to
\begin{eqnarray}\label{eq:rhonP}
\rho^{a (0)}_{n {\bf k}}(P)\simeq \int | \tilde\psi_n({\bf P})|^2 \frac{d\Omega_{\bf p}}{4\pi}
\equiv\rho_n(P),
\end{eqnarray}
where 
\begin{eqnarray}\label{eqn:momspacewf}
 \tilde\psi_n({\bf P})= (2\pi)^{-3/2}\int e^{-i{\bf P}\cdot {\bf r}}\psi_{n}({\bf r})\,d^3r,
\end{eqnarray}
is the momentum space wavefunction, and $\rho_n$ is the spherically averaged electron momentum density of the orbital $n$.\footnote{The density is normalised by $\int \rho_n(P)\,4\pi P^2dP=1$.}
The $\gamma$-spectrum [cf. \eqn{eqn:spectra_sumf}] then takes the form
\begin{eqnarray}\label{eqn:pwspectra}
w(\epsilon)= \frac{1}{c}\int_{2|\epsilon|/c}^{\infty} \rho (P)\, 4\pi P dP,
\end{eqnarray}
where $\rho (P)=\sum_n\rho_n(P)$ is the total electron momentum density of the occupied orbitals in the molecule.
It is related to the cross section measured using the electron-momentum spectroscopy technique~\cite{emsreview}.

\subsection{Direct annihilation rate from $\gamma$-spectra}
The electron-positron direct annihilation rate in a gas of density $n$ is commonly parameterized through the dimensionless quantity $Z_{\rm eff}$, defined as the ratio of the observed annihilation rate $\lambda$ to the rate of free electron-positron annihilation
$Z_{\rm eff}\equiv \lambda/\pi r_0^2cn$~\cite{Fraser,pomeranchuk}.
By definition, $Z_{\rm eff}$ quantifies the \emph{effective number of electrons} per target atom or molecule, with which the positron can annihilate. 
In general, it is  different from the true number of electrons $Z$ owing to the effects of, e.g., the nuclear repulsion and positron-electron interactions, which can act to suppress or enhance the electron density in the vicinity of the positron. For most molecules, $Z_{\rm eff}$ is also enhanced
by the vibrational Feshbach resonances~\cite{RevModPhys.82.2557}.
The ratio $Z_{\rm eff}/Z$ therefore gives a measure of the strength of such effects.
$Z_{\rm eff}$ can be calculated in terms of the Doppler spectrum, and hence annihilation momentum density, as
\begin{eqnarray}\label{eqn:zeffspec}
Z_{\rm eff}=\int_{-\infty}^{\infty} w_{\bf k}(\epsilon) d\epsilon= \int_0^{\infty} \rho^{a}_{\bf k}(P)\,4\pi P^2dP.
\end{eqnarray} 
For a plane-wave positron, in the independent-particle-model approximation, it reduces to
$Z_{\rm eff}=Z$, the number of electrons in the molecule.

\section{Gamma spectra from the density functional theory calculations}\label{sec:dft}
In this section we present the results of the calculations of the $\gamma$-spectra for a selection of molecules. 
They were performed assuming a plane-wave positron, and were based on electron-momentum distributions, Eqs.~(\ref{eq:rhonP})--(\ref{eqn:pwspectra}).
Specifically, we used the DFT-based Becke-3-parameter Lee-Yang-Parr (B3LYP) model~\cite{becke,PhysRevB.37.785}, with the Godbout polarised triple-$\zeta$ valence (TZVP) basis set~\cite{TZVP},  which is known to produce good agreement with the experimental measurements for the molecular orbital momentum profiles~\cite{fengemds}. 
These calculations were performed using the computational chemistry package {\tt Gaussian 03}~\cite{g03}. For C$_3$H$_8$ and C$_6$H$_6$ we used the BP86 functional in {\tt GAMESS} \cite{GAMESS}. The calculations employed finite
cut-off momentum values in calculating the spectra from
\eqn{eqn:pwspectra} (in most cases, $P_{\rm max}=3$~a.u.). When possible, tails of the form $AP^{-\xi} $ were added to the orbital momentum densities to obtain more accurate spectra. In most cases this did not change their widths by more than 5\%.

\begin{figure}[t!]
\centering
\includegraphics*[width=0.85\textwidth]{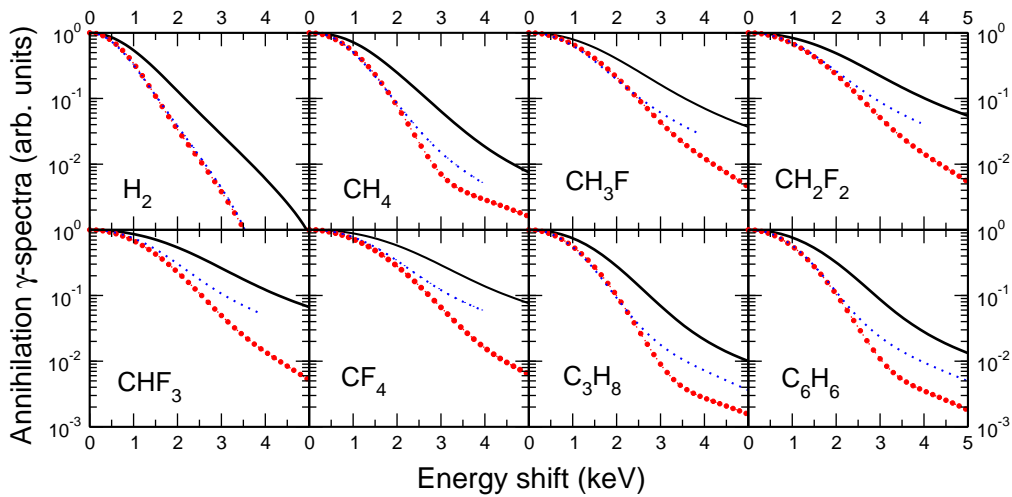}
\vspace{-12pt}
\caption{Annihilation $\gamma$-spectra for a selection of molecules: {\bf \full }, DFT calculation (neglecting the contributions of core orbitals, i.e., C $1s$ and F $1s$); {\color{blue} \dotted }, DFT spectra scaled as
described at the end of section~\ref{subsec:H2gamma};
{\color{red} $\bullet $}, experiment (two-Gaussian fit)~\cite{PhysRevA.55.3586}.\label{fig:othermols} }
\end{figure} 

Figure\,\ref{fig:othermols} shows the results of our plane-wave-positron DFT calculation for the annihilation $\gamma$-spectra of hydrogen, the fluorosubstituted methanes, and propane and benzene.
Also shown are the experimental results of Iwata \emph{et al.}~\cite{PhysRevA.55.3586} obtained assuming that the intrinsic annihilation line shapes can be described by the two-Gaussian form,
\begin{eqnarray}\label{eqn:2gauss}
w(\epsilon)\propto \exp \left(-\epsilon ^2/a^2\right) +A\exp \left(-\epsilon ^2/b^2\right),
\end{eqnarray}
where $A$, $a$ and $b$ are constants obtained by deconvoluting the intrinsic spectra from the measured line shapes. Their values for an extensive range of molecules can be found in Ref.~\cite{PhysRevA.55.3586}.

\begin{table}[ht!]
 \caption{ \label{table:fwhmothermols}Full widths at half-maximum (keV) of the calculated and measured $\gamma$-spectra for molecules.}
\begin{indented}
\item[]\begin{tabular}{lc@{\hspace{6pt}}c@{\hspace{6pt}}c@{\hspace{6pt}}c@{\hspace{6pt}}c@{\hspace{6pt}}c@{\hspace{6pt}}c@{\hspace{6pt}}c@{\hspace{6pt}}c@{\hspace{6pt}}c}
\br	
& 			H$_2$ 	& CH$_4$ 	& CH$_3$F		&CH$_2$F$_2$ 	& CHF$_3$	& CF$_4$		
&C$_2$H$_6$ 	
&C$_3$H$_8$ 
& C$_4$H$_{10}$ 
& C$_6$H$_6$ 	 \\
\mr
DFT$^a$
& 2.14		& 2.81		& 3.40 			& 3.81 			& 4.08 	& 4.29	
& 3.05		
& 2.99			
& 3.12
& 3.14			\\
Experiment$^b$
& 1.59		& 2.06		& 2.64			& 2.72			& 2.71		& 2.92		
& 2.15			
& 2.18			
& 2.25			
& 2.20\\
\br
\end{tabular}
\item[] $^{\rm a}$ B3LYP or BP86 calculation of the electron momentum density (i.e., the plane-wave positron approximation) without the tails. Core-electron contributions are omitted, except for C$_2$H$_6$ and C$_4$H$_{10}$.	
\item[] $^{\rm b}$ Experimental values from the two-Gaussian fit of Ref.~\cite{PhysRevA.55.3586}.
 \end{indented}
 \end{table}

\begin{figure}[ht!]
\centering
\includegraphics*[width=0.5\textwidth]{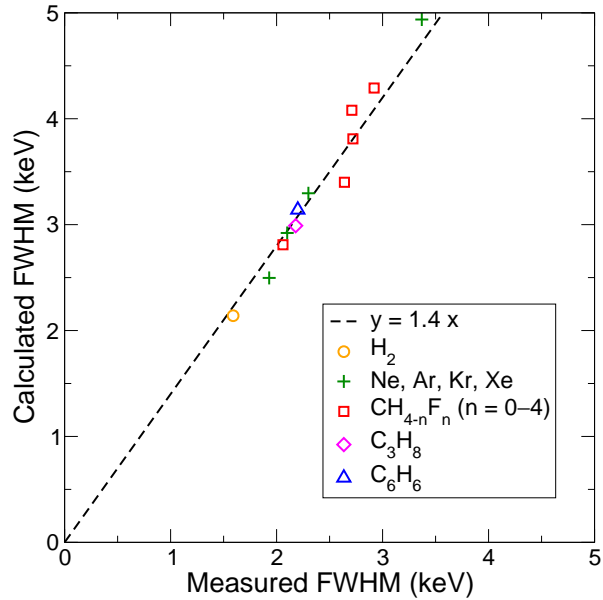}
\caption{FWHM values of the $\gamma$-spectra calculated in the plane-wave-positron approximation vs.~experimental values from Ref.~\cite{PhysRevA.55.3586}, for a range of molecules and noble-gas atoms. The dashed line is a linear relation $y=1.4\,x$.}
\label{fig:FWHM} 
\end{figure}

The values of the calculated and measured full-widths at half-maximum (FWHM) are given in \tab{table:fwhmothermols}.
For all of the molecules studied, the calculation systematically overestimates the measured line widths, although the main chemical trends are reproduced. This can also be seen
in \fig{fig:FWHM}, where we plot the calculated FWHM versus the measured linewidths for all of the molecules discussed, and for the noble gases Ne, Ar, Kr and Xe that were calculated using the Hartree-Fock method~\cite{wangnobles}.\footnote{Note that for the noble gas atoms, many-body theory calculations are much superior to the plane-wave-positron HF method shown here, and produce $\gamma$-spectra in excellent agreement with experiment~\cite{0953-4075-39-7-008, DGG_innershells}.} 
The graph shows that the FWHM values from plane-wave positron approximation are greater than the experimental values by about 40\%.

The main origin of the discrepancy between these calculations and experiment is the neglect of the effect of the positron-molecule potential, and in particular, the positron repulsion from the nuclei, in the calculation. It leads to an overestimate of the electron-positron wavefunction overlap at small positron-nuclear separations, and therefore, of the large momentum contributions to the annihilation momentum density, and of the large energy shifts in the $\gamma$-spectra.

One should expect that the proper inclusion of the positron-nuclear repulsion will lead to a narrowing of the spectra, bringing them more into better agreement with experiment. 
The remainder of the paper outlines a scheme which facilitates this inclusion.
\section{Effect of positron-atom interaction}\label{sec:LCAO}
\subsection{Electron momentum density: linear combination of atomic orbitals}\label{sec:lcao}
To account for the effect of the nuclear repulsion and electron-positron correlations, it is instructive to derive the electron momentum density for a given molecular orbital using a multicentre linear combination of atomic orbitals (LCAO). To do this we follow the approach of Ref.~\cite{Kaijser197737}.

First, consider a generic molecular orbital $\psi({\bf r})$ as a linear combination of the atomic orbitals $\psi _i$ that constitute the molecule,
\begin{equation}\label{eqn:lcao}
\psi ({\bf r})=\sum_i c_i\psi_i({\bf r}-{\bf R}_i),
\end{equation}
where $c_i$ are the expansion coefficients,\footnote{In general the expansion coefficients are governed by the point-group symmetry of the molecule, and by requirements of normalization and energy minimization.}  and ${\bf R}_i$ are the coordinates of the atomic centres.\footnote{In the Born-Oppenheimer approximation the set of the ${\bf R}_i$ are external parameters as opposed to dynamical variables.}

Assuming the atomic orbitals can be written in the central-field form $\psi_{nlm}({\bf r})=R_{nl}(r)Y_{lm}(\hat {\bf r})$, where $R_{nl}$ is the radial function and $Y_{lm}$ is a spherical harmonic, the
corresponding momentum space wavefunction (\ref{eqn:momspacewf}) is
\begin{eqnarray}
\tilde{\psi}_{nlm}({\bf P})=(-i)^l \tilde R_{nl}(P)Y_{lm}(\hat{\bf P}),
\end{eqnarray}
where $\tilde R_{nl}(P)=\sqrt{\frac{2}{\pi}} \int_0^\infty j_l(Pr) R_{nl}(r) r^2 dr$.
The spherically averaged momentum density for this atomic orbital, \eqn{eq:rhonP},  is then
given by
\begin{equation}\label{eqn:rhoR}
\rho_{nl}(P)=  \tilde R_{nl}^2(P)/4\pi .
\end{equation}
Hence, the momentum-space wavefunction can then be written as
\begin{eqnarray}
\tilde \psi _i({\bf P})=(-i)^{l_i}\sqrt{4\pi\rho_i(P)}Y_{l_im_i}(\hat{\bf P}),
\end{eqnarray}
where $i$ labels the specific atomic orbital in the linear combination (\ref{eqn:lcao}) and implicitly includes the labels for the quantum numbers $n$ and $l$.
The momentum-space molecular orbital can then be expressed as follows:
\begin{eqnarray}
\tilde \psi({\bf P})&=&(2\pi)^{-3/2}\int e^{-i{\bf P}\cdot {\bf r}}\psi({\bf r})\,d^3r, \\
&=&\sum_i c_i e^{-i{\bf P}\cdot {\bf R}_i}{\tilde \psi}_i({\bf P}),
\end{eqnarray}
and the corresponding momentum density $\rho (P)$ is easily shown to be
\begin{eqnarray}\label{eq:rhomol}
\fl\rho(P)=\sum_{i,j} (-i)^{l_j-l_i}  c^*_{i} c_{j}\sqrt{\rho_i(P)\rho_j(P)}
\int e^{i{\bf P}\cdot({\bf R}_i-{\bf R}_{j})} Y^*_{l_im_i}(\hat{\bf P})Y_{l_jm_j}(\hat{\bf P})\,d\Omega_{\bf P}.
\end{eqnarray}
It is convenient to separate the diagonal and off-diagonal parts in the above double sum, and
to split the latter into sums involving single-centre and two-centre terms:
\begin{equation}\label{eqn:rhomolall}
\rho(P)= \sum_{i}{|c_i|^2\rho_i(P)}+ \mathop{\sum_{ {\bf R}_i={\bf R}_j }}_{i, \,j>i}{\rho^{\rm I}_{ij}(P)} + \mathop{\sum_{ {\bf R}_i\neq{\bf R}_j }}_{i, \,j\neq i}{\rho^{\rm II}_{ij}(P)}.
\end{equation}
In this equation, the second term on the right-hand-side is a sum of the products of wavefunctions centred on the same atom. They satisfy the usual orthogonality relations and take the form
\begin{eqnarray}\label{eqn:rhomol1}
{\rho^{\rm I}_{ij}(P)}=(c_i^* c_j + c_ic_j^*) \delta_{l_il_j}\delta_{m_im_j} \sqrt{\rho_i(P)\rho_j(P)}.
\end{eqnarray}
The third term on the right-hand-side of \eqn{eqn:rhomolall} is a sum over the product of wavefunctions centred on different atomic sites.
Resolving the plane wave and expressing the integrals of three spherical harmonics in terms of $3j$ symbols (see e.g., \cite{varshalovich}), it can be written as
\begin{eqnarray}\label{eqn:rhomolgen}
\fl\rho^{\rm II}_{ij}(P)= \sqrt{4\pi}\, c^*_{i} c_{j}\sqrt {\rho_i(P)\rho_j(P)} (-1)^{m_i}
\sum_{\lambda, \mu} (-1)^{(l_i-l_j+\lambda)/2}\, j_{\lambda}(PR_{ij})
Y^*_{\lambda \mu}(\hat{{\bf R}}_{ij})  \nonumber \\ 
\times
\sqrt{[l_i][l_j][\lambda]}\left(
  \begin{array}{ c c c }
     l_i 			& \lambda 		& l_j \\
     -m_i		 	& \mu	 	& m_j
  \end{array} \right)
  \left(
  \begin{array}{ c c c }
     l_i			& \lambda  		& l_j \\
     0		 	& 0				& 0
  \end{array} \right),
\end{eqnarray}
where ${\bf R}_{ij}={\bf R}_i-{\bf R}_j$ and we use the notation $[l]\equiv(2l+1)$.

To summarize, in the LCAO approach the \emph{molecular} orbital momentum density is determined by the products of the relevant \emph{atomic} momentum densities,
\begin{eqnarray}\label{eqn:mollcao}
\rho(P)= \sum_{i,j\geqslant i}C_{ij}(c_i,P,{\bf R}_{ij}) \sqrt{\rho_i(P)\rho_j(P)},
\end{eqnarray}
where $C_{ij}$ are the prefactors given in Eqs.~(\ref{eqn:rhomolall})--(\ref{eqn:rhomolgen}).
This quantity represents the annihilation momentum density for the corresponding molecular orbital in the low-energy plane-wave-positron approximation. A more accurate molecular annihilation spectrum will be obtained by replacing the atomic electron momentum densities in \eqn{eqn:mollcao} by the corresponding annihilation momentum densities.

\subsection{Atomic correction factors}\label{sec:adjustmenttheory}
To include the effects of the positron-atom interactions in the molecular annihilation momentum density, the plane-wave \emph{atomic} electron momentum densities $\rho_i(P)$ that appear in \eqn{eqn:mollcao} are multiplied by the \emph{atomic correction factor}, defined as
\begin{eqnarray}\label{eq:GP}
G_i(P)\equiv\frac{\rho_i^a(P)}{\rho_{i}(P)},
\end{eqnarray}
i.e., as the ratio of the true annihilation momentum density of the atomic orbital $i$ to its
plane-wave independent-particle approximation.

\begin{figure}[ht!]
\centering
\includegraphics*[width=0.8\textwidth]{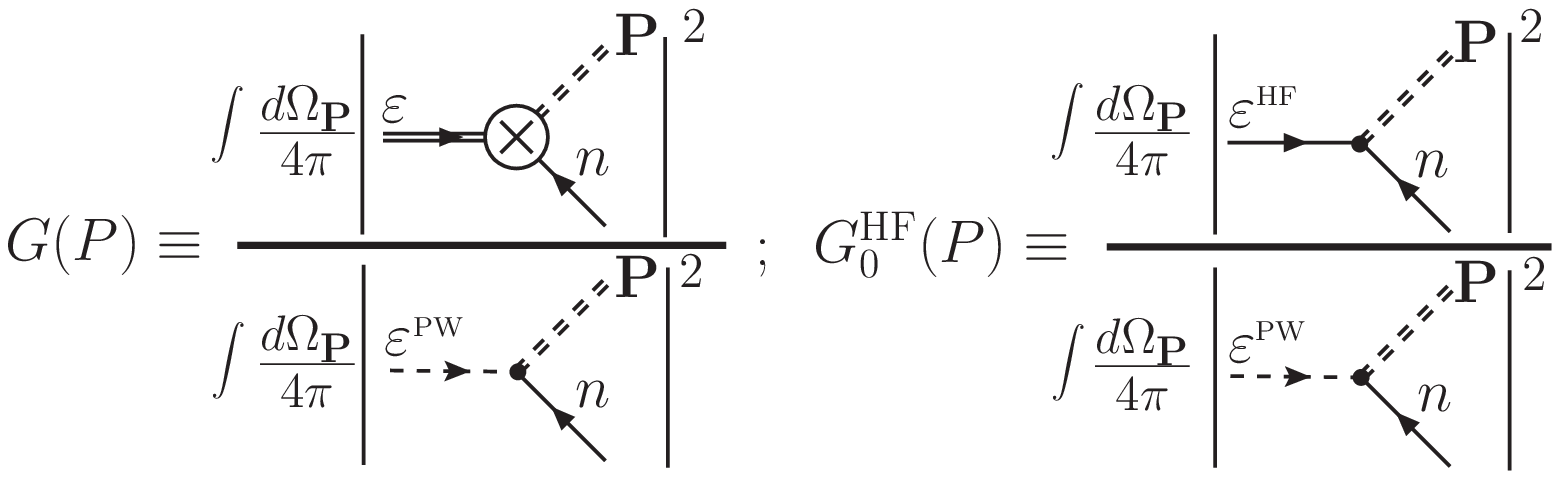}\\
~~~~~(a)~~~~~~~~~~~~~~~~~~~~~~~~~~~~~~~~~~~~~~~~~~(b)
\caption[Definition of atomic correction factors]{Atomic correction factors: (a) the `exact' atomic correction factor for atomic orbital $n$, defined as $G(P)\equiv\rho_n^{a}(P)/\rho_n(P)$, where $\rho_n^a$ is calculated using the positron Dyson orbital and the full vertex (see \fig{fig:anndiagrams}); (b) the simpler Hartree-Fock correction factor defined as $G_0^{\rm HF}(P)\equiv\rho_n^{a, \rm HF}(P)/\rho_n(P)$, where $\rho_n^{a, \rm HF}(P)$ is calculated with a Hartree-Fock positron wavefunction and the zeroth-order annihilation vertex.}
\label{fig:g_factordef}
\end{figure}

For atomic hydrogen, the correction factors can be calculated accurately from many-body theory (see \fig{fig:g_factordef}\,(a) for a diagrammatic definition).
They therefore include the effect of electron-positron and positron-atom correlations, in addition to that of the nuclear repulsion.
A simpler correction factor that can be used for any atom and still include the effect of the atomic potential and nuclear repulsion can be obtained  using a Hartree-Fock (HF) positron function with the zeroth-order (independent-particle) annihilation vertex (see \fig{fig:g_factordef}\,(b)).
Note that the effect of the short-range electron-positron correlations are not included in this correction factor.

\begin{figure}[ht!]
\centering
\includegraphics*[width=0.6\textwidth]{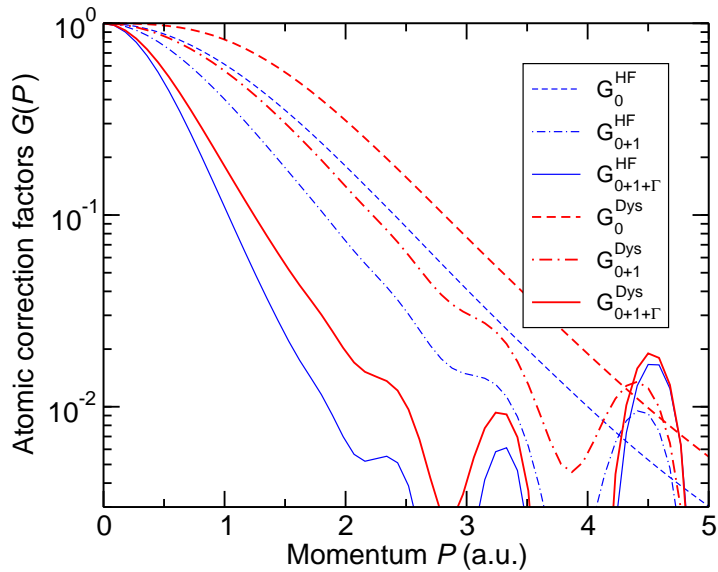}
\vspace{-12pt}
\caption[Atomic correction factors for hydrogen]
{Momentum-dependent atomic correction factors for hydrogen calculated
with the Hartree-Fock (i.e., static field) and Dyson positron orbitals using different approximations for the annihilation vertex. All correction factors are normalised to unity at $P=0$. The oscillations in the tails are artefacts arising from computational ``noise'' in the  many-body theory calculations.}
\label{fig:g_factorH}
\end{figure}

\begin{figure}[t!]
\centering
\includegraphics*[width=0.6\textwidth]{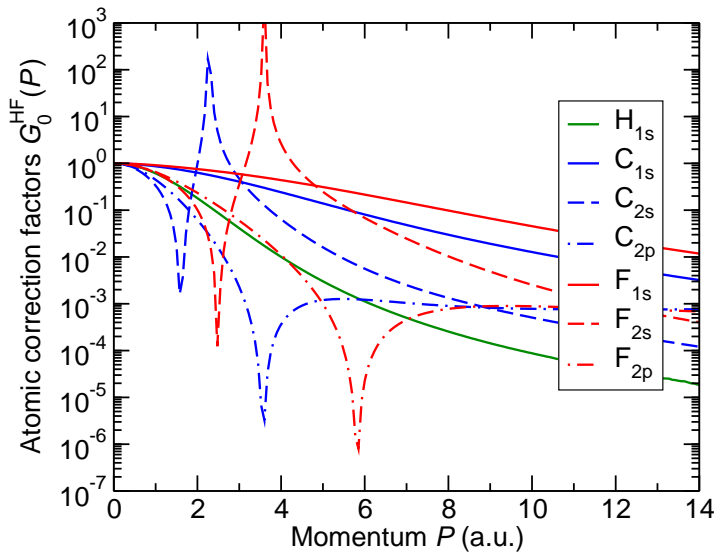}
\vspace{-12pt}
\caption[Atomic correction factors for carbon and fluorine]
{Momentum-dependent atomic correction factors for the ground state orbitals of hydrogen, carbon and fluorine calculated using the Hartree-Fock positron wavefunction with the zeroth-order annihilation vertex.
All correction factors are normalised to unity at $P=0$.}
\label{fig:g_factorCF}
\end{figure}

We will apply this procedure to calculate the $\gamma$-spectra of H$_2$, CH$_4$ and CF$_4$.
Figures \ref{fig:g_factorH} and \ref{fig:g_factorCF} show the atomic correction factors for hydrogen, carbon and fluorine, calculated for the $s$-wave positron wavefunction which dominates at the low (room-temperature) momentum $k=0.04$~a.u. To highlight the effect that these factors have on the \emph{shape} of
$\gamma $-spectra, they are shown normalised to unity at $P=0$.

Figure \ref{fig:g_factorH} shows the correction factor for hydrogen calculated using different approximations to the annihilation vertex and positron wavefunction. The effect of the nuclear repulsion is most easily seen through the zeroth-order factor $G^{\rm HF}_0$. 
The shape of this factor is mainly governed by the nuclear repulsion. It is evident that the effect of the nuclear repulsion on the positron is to suppress the high-momentum regions of the momentum density. Further suppression results from inclusion of the additional correlations. 
One can also see that for a given approximation to the positron wavefunction, adding the first-order and higher-order ($\Gamma$-block) correlation corrections to the vertex increases the degree of suppression of high momentum components to the annihilation momentum density.  
Physically, this is explained by noting that the corrections to the vertex involve positron
annihilation with spatially diffuse virtual electrons. 
On the other hand, using the correlated positron Dyson orbital instead of the Hartree-Fock
(i.e., static field) positron wavefunction increases the high-momentum component for a given
approximation for the vertex. Again this is what one should expect physically, since the inclusion of the attractive correlational potential accelerates the positron towards the atom and increases
the momentum of the annihilating electron-positron pair. Regarding the $\gamma$-spectrum shape therefore, there is some compensation between including the correlation corrections to the vertex and increasing the accuracy of the positron wavefunction. However, the dominant effect of $G(P)$ is always the suppression of high momenta.

For the carbon and fluorine atoms, we have only the zeroth-order Hartree-Fock correction factors at our disposal\footnote{The many-body theory cannot be applied directly to open-shell atoms, but the static-field effect is the dominant feature for higher-$Z$ atoms and core orbitals.} and these are shown for the ground states orbitals in \fig{fig:g_factorCF}.
Note that the correction factors for the $2s$ orbitals show singularities arising from the nodes in the Hartree-Fock wavefunctions (which result in the zeros of the corresponding electron momentum densities $\rho _{2s}(P)$).
Nevertheless, the suppression of high momentum contributions to the spectra is clear.
Moreover, when evaluating the $\gamma$-spectra in the LCAO approach we circumvent the problem of the singularities in the correction factors by simply replacing the atomic electron momentum densities with the annihilation momentum densities, i.e., we calculate $\rho^a_i(P)$ directly rather than $\rho_i(P)G_i(P)$.  
For carbon and fluorine, the positrons predominatly annihilate on the valence electrons. 
The magnitude of the correction factor for the valence electrons is therefore substantially greater than for the core electrons. 

Finally, although for hydrogen the correction factors are sensitive to the positron-atom and positron-electron correlations, for heavier atoms most of the suppression at high momenta is due
to the effect of the nuclear repulsion. Hence, for such atoms the HF correction factor will describe the true suppression more accurately than it does for hydrogen.
This point will be discussed further in \secref{sec:results}.

\section{Gamma spectra with atomic correction factors}\label{sec:results}
\subsection{Molecular hydrogen}\label{sec:hydrogen}
\nopagebreak
\subsubsection{Gamma spectrum of H$_2$}\label{subsec:H2gamma}
In the LCAO approximation, the ground state molecular orbital $\psi _{\sigma_g}({\bf r})$ can be written a linear combination of two hydrogenic ${1s}$ orbitals, each centred on one of the protons,
\begin{equation}\label{eqn:h2lcaomo}
\psi_{\sigma_g}({\bf r})= \frac{1}{\sqrt{2(1+S)}} \left[ \psi_{1s}({\bf r}-{\bf R}_1) + \psi_{1s}({\bf r}-{\bf R}_2)\right],
\end{equation}
where $\psi_{1s}({\bf r})=(\pi \alpha ^3)^{-1/2}e^{-\alpha r}$ and $S$ is the overlap integral $\int \psi _{1s}({\bf r}-{\bf R}_1)\psi_{1s}({\bf r}-{\bf R}_2)d^3r$. It can be calculated analytically using prolate spheroidal coordinates as $S=e^{-\alpha R}\left[1+\alpha R+\frac{1}{3}(\alpha R)^2\right]$, where $R$ is the internuclear distance~\cite{slater}.
A variational calculation gives $\alpha\approx 1.19$ for $R=1.38$~a.u., and therefore, $S\approx 0.69$~\cite{DGG_thesis}.

\begin{figure}[t!]
\centering
\includegraphics*[width=0.6\textwidth]{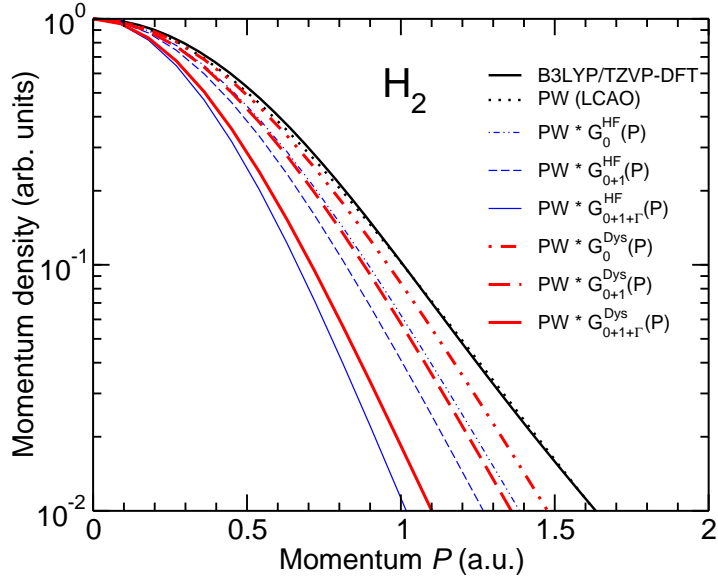}
\caption{Hydrogen molecule electron momentum densities calculated using the B3LYP/TZVP DFT (black solid curve) and LCAO, \eqn{eqn:h2lcaomd} (dots), and annihilation momentum densities obtained from \eqn{eqn:h2lcaomd} by multiplying the ``plane-wave'' (PW) atomic density
$\rho_{1s}(P)$ by the correction factors $G(P)$ obtained from many-body-theory calculations
using various approximations to the annihilation vertex and positron wavefunction (see legend). All densities are normalised to unity at $P=0$.}
\label{fig:h2md}
\end{figure}

Substituting \eqn{eqn:h2lcaomo} into \eqn{eqn:rhomolall} gives the molecular orbital momentum density
\begin{equation}\label{eqn:h2lcaomd}
\rho_{\sigma_g}(P) = \frac{1}{1+S}\, \rho_{1s}(P) \left[1 + j_0(PR)\right],
\end{equation}
where $\rho_{1s}(P)$ is the momentum density of the hydrogenic $1s$ orbital,
\begin{eqnarray}\label{eq:rho1s}
\rho_{1s}(P)=\frac{8\alpha^5}{\pi^2(\alpha^2+P^2)^4}.
\end{eqnarray}
Figure~\ref{fig:h2md} shows that the LCAO molecular orbital density given by \eqn{eqn:h2lcaomd} is very close to that from B3LYP/TZVP density functional theory calculation.

Figure~\ref{fig:h2md} also shows the hydrogen molecule annihilation momentum densities 
obtained from \eqn{eqn:h2lcaomd} by scaling the atomic momentum density with the correction factors $G(P)$ obtained from many-body-theory calculations using various approximations to the annihilation vertex and positron wavefunction (cf. \fig{fig:g_factorH}), i.e., by substitution $\rho_{1s}(P)\to G(P)\rho_{1s}(P)$. To emphasise the effect of the correction factors on the annihilation momentum density, all densities are normalized to unity at $P=0$. As expected, the use of correction factors leads to significant narrowing of the momentum ditribution.

\begin{figure}[t!]
\centering
\includegraphics*[width=0.6\textwidth]{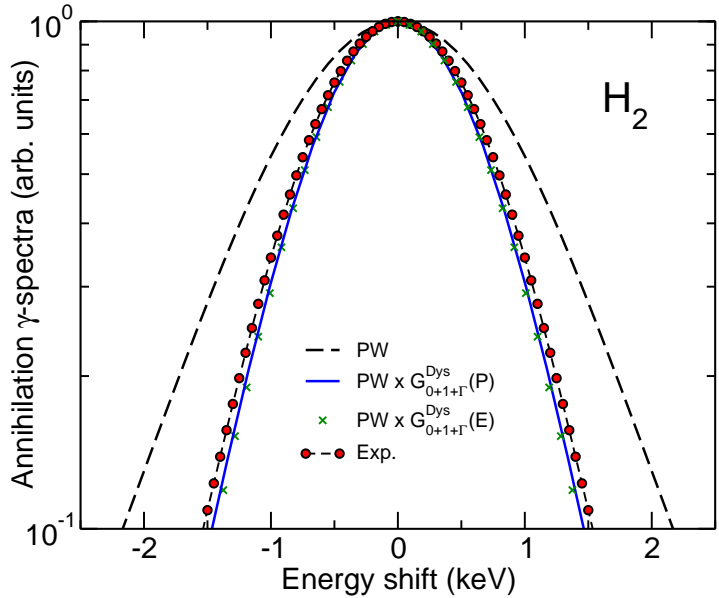}
\caption{Annihilation $\gamma$-spectra of H$_2$ calculated using LCAO in the plane-wave-positron approximation, i.e., using the density from \eqn{eqn:h2lcaomd} (long-dashed line), and
incorporating the `exact' momentum-dependent atomic correction factor $G^{\rm Dys}_{0+1+\Gamma}(P)$ (solid line) and the energy-shift-dependent correction factor $G_\gamma (\epsilon)$
(crosses, see text); solid circles show the measured spectrum, as represented by the two-Gaussian fit, Ref.~\cite{PhysRevA.55.3586}. All spectra are normalised to unity at $\epsilon =0$.}
\label{fig:h2asnormed}
\end{figure}

The corresponding annihilation $\gamma$-spectra are shown in \fig{fig:h2asnormed}, where they are compared with the experimental result of Iwata \emph{et al.}~\cite{PhysRevA.55.3586}. 
The FWHM values of the $\gamma$-spectra are given in table~\ref{table:h2fwhm}. 
The spectra obtained from the B3LYP/TZVP and LCAO electron momentum densities are practically indistinguishable, and we show only the latter in the figure.
This `low-energy plane-wave positron' calculation gives the FWHM value about 35\% greater than experiment. As discussed above, this overestimation of the high-momentum contribution is largely
 due to the neglect of the positron-nuclear repulsion.
Introducing the effects of the nuclear repulsion through the zeroth-order vertex Hartree-Fock correction factor $G_0^{\rm HF}$, narrows the spectrum, but it remains $\approx 18\%$ broader than experiment. 
Using the most accurate, `exact' atomic adjustment factor $G_{0+1+\Gamma }^{\rm Dys}(P)$ gives the FWHM that is in very good agreement (within $6\%$) of the measured value. 
This shows that for H$_2$, the inclusion of the nuclear repulsion provides only 50\% of the required narrowing to bring the plane-wave calculation into agreement with experiment, with the remaining 50\% being due to the electron-positron correlations. However, as the nuclear potential becomes stronger in molecules containing heavier atoms, the effect of the potential on the linewidth is expected to become the principal cause of the narrowing, overriding the effect of the correlations.

\begin{table}[hpt!]
\caption{\label{table:h2fwhm} FWHM of the annihilation $\gamma$-spectra and annihilation rate parameter $Z_{\rm eff}$ for H$_2$ calculated in the plane-wave positron approximation (B3LYP/TZVP DFT and LCAO) and using the
atomic correction factors in different approximations, and results of other theoretical calculations and experiments. }
\begin{indented}
\item[]\begin{tabular}{@{}lcc}
\br
Method 	& {FWHM} 	& $Z_{\rm eff}$ \\
		& (keV) 		&\\
\mr
B3LYP/TZ-DFT (plane-wave)$^{\rm a}$					& 2.14		& 2.00			\\			
LCAO (plane-wave)$^{\rm a}$							& 2.14		& 2.00			\\
LCAO$\times G_0^{\rm HF}(P)$$^{\rm a}$				& 1.88		& 0.78			\\
LCAO$\times G_{0+1}^{\rm HF}(P)$$^{\rm a}$				& 1.73		& 1.98			\\
LCAO$\times G_{0+1+\Gamma}^{\rm HF}(P)$$^{\rm a}$		& 1.40		& 4.59			\\
LCAO$\times G_0^{\rm Dys}(P)$$^{\rm a}$ 				& 2.01		& 2.62			\\
LCAO$\times G_{0+1}^{\rm Dys}(P)$$^{\rm a}$				& 1.85		& 6.54			\\
LCAO$\times G_{0+1+\Gamma }^{\rm Dys}(P)$$^{\rm a}$	& 1.50		& 13.9			\\[0.5ex]
\mr
Kohn variational$^{\rm b}$							& --			& 10.3			\\
Kohn variational$^{\rm c}$							& --			&  9.8			\\
Kohn variational$^{\rm d}$							& --			& 12.6			\\	
Stochastic variational$^{\rm e}$						& --			& 15.7			\\
Molecular $R$-matrix$^{\rm f}$							& --			& 10.4			\\
\hline\\[-2ex]
Expt.$^{\rm g}$										& 1.59		& --				\\
Expt.$^{\rm h}$										& --			& 14.7			\\
Expt.$^{\rm i}$										& --			& 16.0			\\
Expt.$^{\rm j}$										& --			& 14.6			\\[0.5ex]
\br
 \end{tabular}
\item[] $^{\rm a}$ This study, $R=1.4$~a.u., $k=0.04$~a.u.
\item[] $^{\rm b}$ `Method of models' calculation from Ref. \cite{armourh2_0}.
\item[] $^{\rm c}$ Calculation for $R$=1.4\au, $k=0.04$\au, Ref. \cite{armourh2_1}.
\item[] $^{\rm d}$ `Method of models' calculation from Ref. \cite{armourh2_new}.	
\item[] $^{\rm e}$ Confined variational method calculation, Ref.~\cite{PhysRevLett.103.223202}.
\item[] $^{\rm f}$ Molecular $R$-matrix with pseudo-states, Ref.~\cite{ruizhangh2}.
\item[] $^{\rm g}$ Experiment, room-temperature positrons, Ref.~\cite{PhysRevA.55.3586}.			
\item[] $^{\rm h}$ Experiment, room-temperature positrons, Ref.~\cite{PhysRevA.20.347}.
\item[] $^{\rm i}$ Experiment, room-temperature positrons, Ref.~\cite{charlton_h2}.
\item[] $^{\rm j}$ Experiment, room-temperature positrons, Ref.~\cite{larrichia_h2}.
\end{indented}
 \end{table}

In addition, we can test a simpler method in which the correction factor is applied directly to the Doppler-shift spectrum rather than in the two-$\gamma$ momentum density.
The method is suggested by an observation that the momentum densities and the $\gamma $-spectra in figures \ref{fig:h2md} and \ref{fig:h2asnormed} are approximately Gaussian (i.e., parabolic in the semilogarithmic plots).
If one assumes that both the electron momentum and the true annihilation momentum \emph{atomic} densities are Gaussians, then then the atomic correction factor $G(P)$, \eqn{eq:GP}, will also be a Gaussian. It is easy to see from Eqs.~(\ref{eqn:spectra_sumf}) and (\ref{eqn:pwspectra}) that
in this case the corresponding annihilation spectra are also Gaussian. Their widths are proportional to those of the momentum densities (scaled by $c/2$). Denoting the FWHM values of the true annihilation spectrum and that found using the electron momentum density by $\sigma _a$ and $\sigma _m$, we can introduce the atomic $\gamma$-spectrum correction factor $G_\gamma (\epsilon)\equiv w(\epsilon)/w_m(\epsilon)$, which will also be a Gaussian, of width $\sigma_G=\left(\sigma^{-2}_a-\sigma^{-2}_m \right)^{-1/2}$.

This means that if the $\gamma$-spectrum of \emph{molecular} hydrogen calculated using the electron momentum density (i.e., plane-wave positron) is close to a Gaussian, of width $\sigma_m^{{\rm H}_2}$, then one can approximate the true annihilation $\gamma$-spectrum of H$_2$ by $w^{{\rm H}_2}(\epsilon)\approx G_\gamma (\epsilon)w_m^{{\rm H}_2}(\epsilon)$. 
The latter will also be a Gaussian with the width
\begin{eqnarray}\label{eq:sigmamG}
\sigma_a^{{\rm H}_2}=\left[ \frac{1}{(\sigma_m^{{\rm H}_2})^2}+\frac{1}{\sigma^2_G}\right]^{-1/2}
=\frac{\sigma_m^{{\rm H}_2}}{\sqrt{1+(\sigma_m^{{\rm H}_2}/\sigma_G)^2}}.
\end{eqnarray}
Figure \ref{fig:h2asnormed} shows that adjusting the
the spectrum by the Doppler-shift-dependent correction factor $G_\gamma (\epsilon)$ for H$_2$ produces a spectrum in good agreement with the result obtained using $G(P)$ and with experiment.

Estimating the width of $G_\gamma (\epsilon)$ from the best calculation shown in \fig{fig:g_factorCF} ($G_{0+1+\Gamma}^{\rm Dys}$) gives
$\sigma _G=2.10$~keV. Using this value together with the plane-wave-positron
width $\sigma _m^{{\rm H}_2}=2.14$~keV (table~\ref{table:h2fwhm}) in
\eqn{eq:sigmamG} gives the annihilation spectrum width $\sigma _a^{{\rm H}_2}\approx 1.50$~keV, in agreement with the best calculated value in table~\ref{table:h2fwhm}.

This example shows that $\sigma _G\approx \sigma _m$, i.e., that the width of the positron atomic correction factor is close to the that of the electron momentum distribution. In this case the expression under the square root in \eqn{eq:sigmamG} is close to 2, giving $\sigma_a\approx \sigma_m/1.4$. It appears that this relation is quite general, as shown
by the FWHM values for other molecules and atoms in table~\ref{table:fwhmothermols} and \fig{fig:FWHM}. This suggests that to obtain
a closer description of experiment, the spectra calculated neglecting the positron can be modified by simply scaling the energy shift by a factor~1.4. The corresponding spectra are shown in \fig{fig:othermols} by dotted lines. The agreement with experiment is excellent for hydrogen and good for alkanes (except in the wings), with larger discrepancies for molecules containing fluorine atoms.

\subsubsection{$Z_{\rm eff}$ of H$_2$} 
 
The correction factor approach also provides a means of estimating the normalized annihilation rate parameter $Z_{\rm eff}$ for positrons on H$_2$.  Its accurate theoretical calculation has proved to be a difficult problem to which considerable attention has been paid. 
Extensive calculations have been performed by Armour and co-workers, who have applied various versions of the Kohn variational method~\cite{armourh2_0,armourh2_1,armourh2_new}.
However, their best theoretical results still underestimate commonly accepted measured value of 14.6~\cite{larrichia_h2}. Only recently has this problem been effectively solved, with Zhang \emph{et al.} \cite{PhysRevLett.103.223202} performing a stochastic variational method calculation which provided an accurate total
wavefunction of this two-elelctron-one-positron system and yielded a low-energy (thermalized) $Z_{\rm eff}$ of 15.7.

In this work we calculate the $Z_{\rm eff}$ from the $\gamma$-spectrum using \eqn{eqn:zeffspec}.
The results of the B3LYP/TZVP DFT and LCAO calculations using the atomic correction factors are given in \tab{table:h2fwhm}.
As one should expect, the plane-wave positron approximation (both B3LYP/TZVP and LCAO)
gives $Z_{\rm eff}=2$, i.e., the number of electrons. The effect of the repulsive atomic field, implicit in $G_0^{\rm HF}$, reduces the magnitude of $Z_{\rm eff}$ by more than a factor of two relative to the plane-wave calculation, owing to the reduced overlap between the electrons and the incident positron. 
However, the successive inclusion of higher-order correlation effects through the correction factors produces a marked increase in the calculated $Z_{\rm eff}$, with the full many-body theory (Dyson positron orbital, exact vertex) giving $Z_{\rm eff}=13.9$, which is within $5\%$ of the measured value. This level of agreement is partly fortuitous, but it is still remarkable that the calculation which uses atomic correction factors gives a molecular $Z_{\rm eff}$ value in such close agreement with experiment.
The fact that it does gives some \emph{a posteriori} justification for the LCAO correction factor approach and its ability to take appropriate account of the positron-molecule and positron-electron correlations in H$_2$.


\subsection{Gamma spectrum of methane}\label{subsec:CH4}

Methane, CH$_4$, has a tetrahedral structure with a point group $T_d$. 
The four hydrogenic $1s$ orbitals span the reducible representation $\Gamma^{\rm red}_\sigma=A_1\oplus T_2$, where $A_1$ is the fully symmetric one-dimensional representation and $T_2$ is a degenerate three-dimensional representation. 
Denoting the carbon ground state orbitals by $\psi _{nl}$ and the four atomic hydrogen ground-state orbitals by $\sigma_i$, the ground state molecular LCAO of methane are (see e.g.,~\cite{cotton})
\begin{eqnarray}
\psi_{1a_1}({\bf r}) &=& \alpha_1 \psi_{1s}+ \beta_1\psi_{2s}+ \gamma_1 \sigma ^{A_1},\label{eqn:1a1}\\
\psi_{2a_1}({\bf r}) &=& \alpha_2 \psi_{1s}+ \beta_2 \psi_{2s}+ \gamma_2 \sigma ^{A_1},\label{eqn:2a1}\\
\psi _{1t_{2x,y,z}}({\bf r})&=& \delta \psi_{2p_{x,y,z}}+ \gamma_3 \sigma _{x,y,z}^{T_2}\label{eqn:t2},
\end{eqnarray}
where $\alpha$, $\beta$, $\gamma$ and $\delta$ are the expansion coefficients and
the symmetry adapted linear combinations (SALC) of pendant atoms are given by
\begin{eqnarray*}\label{eqn:tdsalcs}
{\sigma}^{A_1}({\bf r})&=& \sigma_1+\sigma_2+\sigma_3+\sigma_4 \label{eqn:t2d_sigma_a1salc},\\
{\sigma}^{T_2}_{x}({\bf r})&=& \sigma_1-\sigma_2-\sigma_3+\sigma_4 ,			\\
{\sigma}^{T_2}_{y}({\bf r})&=& \sigma_1+\sigma_2-\sigma_3-\sigma_4,			\\ 
{\sigma}^{T_2}_{z}({\bf r})&=& \sigma_1-\sigma_2+\sigma_3-\sigma_4 .  \label{eqn:t2d_sigma_t2salc} 
\end{eqnarray*}

The total electron momentum density for methane is given by the sum of the respective individual molecular orbital densities,
\begin{equation}\label{eq:rhoCH4}
\rho (P)= 2\rho_{1a_1}(P) + 2\rho_{2a_1}(P) + 6\rho_{1t_2}(P),
\end{equation}
obtained from Eqs.~(\ref{eqn:1a1})--(\ref{eqn:t2}) and \eqn{eqn:rhomolall}
in terms of the atomic orbital momentum densities, as
\begin{eqnarray}
\fl\rho_{na_1}(P)=\alpha_n^2\rho^{\rm C}_{1s}(P)+\beta_n^2\rho^{\rm C}_{2s}(P)
+ 2\alpha_n\beta_n\sqrt{\rho ^{\rm C}_{1s}(P) \rho^{\rm C}_{2s}(P)}+ 4\gamma_n^2\rho^{\rm H}_{1s}(P) \left[1+3\,j_0(PR_{\rm HH}) \right] \nonumber\\
+ 8j_0(PR_{\rm CH})\gamma_n\sqrt{\rho^{\rm H}_{1s}(P)}\left[\alpha _n\sqrt{\rho^{\rm C}_{1s}(P)}+\beta _n\sqrt{\rho^{\rm C}_{2s}(P)}\right],
\label{eq:rhoa1}\\
\fl\rho_{1t_2}(P)= \delta^2\rho^{\rm C}_{2p}(P)+ 4\gamma_3^2\rho^{\rm H}_{1s}(P)\left[1-j_0(PR_{\rm HH})\right]
+ 8\gamma_3 \delta j_1(PR_{\rm CH})\sqrt{\rho^{\rm C}_{2p}(P)\rho^{\rm H}_{1s}(P)}.\label{eq:rhot2}
\end{eqnarray}
where $R_{\rm CH}$ and $R_{\rm HH}$ are the C--H and H--H bond lengths, respectively.

We construct the molecular orbitals of Eqs.~(\ref{eqn:1a1})--(\ref{eqn:t2}) in two ways. 
In the first case, we use Slater-type atomic orbitals (LCAO-STO) optimized for the molecular environment~\cite{pitzer}. 
In the second case, we use Hartree-Fock atomic orbitals (LCAO-HF).
In both cases the LCAO expansion coefficients are taken to be those of Ref.~\cite{pitzer}, with a C--H distance of $R_{\rm CH}=2.05$~a.u.

\begin{figure}[t!]
\centering
\includegraphics*[width=0.6\textwidth]{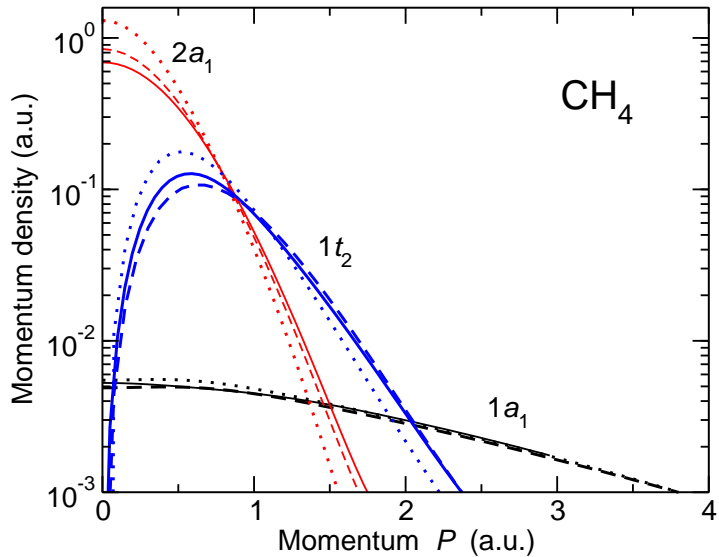}
\caption{Electron momentum densities (i.e., plane-wave positron annihilation momentum densities) for the molecular orbitals of methane: (solid lines) B3LYP/TZVP; (dashed lines) Slater-type orbitals; (dotted lines) Hartree-Fock. }
\label{fig:ch4md} 
\end{figure}

Figure~\ref{fig:ch4md} shows the electron momentum densities calculated using the DFT based B3LYP/TZVP model, and the LCAO result from \eqn{eq:rhoCH4} using the Slater-type and Hartree-Fock orbitals.
For a given molecular orbital, the results of all three calculations are similar.
However,  in general the LCAO-STO calculation is in better agreement with the B3LYP/TZVP calculation than the LCAO-HF one, owing to the fact that the HF orbitals used are purely atomic in nature, i.e., they are in no way adjusted to account for the molecular environment.\footnote{This is in contrast with the treatment of H$_2$ in section~\ref{sec:hydrogen}, where we used hydrogenic orbitals with $\alpha =1.19$ from a variational calculation, rather than those of the H atom ($\alpha =1$).}  
Note that the momentum density of the $1a_1$ orbital (predominantly carbon $1s$ in nature) is much broader than those of the valence $2a_1$ and $1t_2$ orbitals due to the fact that the core electrons have larger characteristic momenta.

To take into account the effect of the positron-atom interactions on the
annihilation momentum density and $\gamma $-spectrum, we replace the
atomic electron momentum densities in Eqs.~(\ref{eq:rhoa1}) and (\ref{eq:rhot2}) by the corresponding annihilation momentum densities calculated using the Hartree-Fock positron wavefunction. These are then used in Eq.~(\ref{eq:rhoCH4}) and the annihilation spectrum is found from Eq.~(\ref{eqn:spectra_sumf}). This procedure is equivalent to the use of atomic correction factors $G_0^{\rm HF}(P)$, which is expected to narrow the spectrum (see \fig{fig:g_factorCF}).

\begin{figure}[ht!]
\centering
\vspace{-10pt}
\includegraphics*[width=0.6\textwidth]{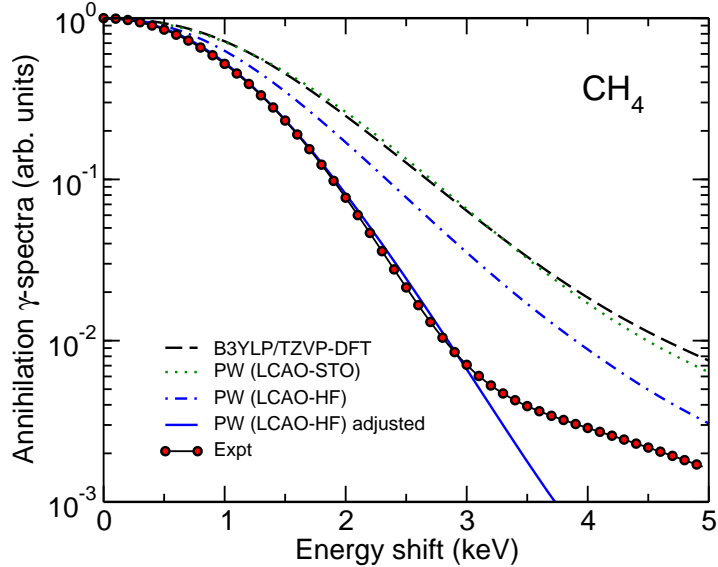}
\caption{Annihilation $\gamma$-spectra of CH$_4$: (dashed line) B3LYP calculation; (dotted line) LCAO-STO; (dot-dashed line) LCAO-HF (all correspond to the low-energy plane-wave positron); (thick solid) LCAO-HF using the atomic annihilation momentum densities; (circles) experiment,
two-Gaussian fit, Ref.~\cite{PhysRevA.55.3586}. All spectra are normalised to unity at $\epsilon =0$.}
\label{fig:ch4as}
\end{figure}

Figure \ref{fig:ch4as} shows the annihilation $\gamma$-spectra of CH$_4$, calculated using the B3LYP/TZVP and LCAO (both STO and HF), and the LCAO-HF results corrected for the positron-atom interactions, as well as the measured spectrum. 
The B3LYP/TZVP  spectrum is significantly broader than the measured one, as seen earlier in \fig{fig:othermols}. 
The fact that the LCAO-STO result is in very good agreement with the B3LYP/TZVP result confirms the applicability of the LCAO approach. The LCAO-HF spectrum calculated in the same low-energy plane-wave positron approximation (i.e., neglecting the positron wavefunction) is narrower
than B3LYP/TZVP, but still much broader than experiment. The overestimation of the width must therefore originate from the neglect of the positron wavefunction. Note that the plane-wave calculation overestimates the spectral FWHM  significantly, even when the $1a_1$ core orbital is neglected (see table~\ref{table:ch4as}).

Correcting the LCAO-HF spectrum by using atomic annihilation momentum densities in Eqs.~(\ref{eq:rhoCH4})--(\ref{eq:rhot2}) produces a $\gamma$-spectrum which is very close to experiment for Doppler shifts smaller than 3~keV (solid line in \fig{fig:ch4as}). The excellent agreement with experiment is somewhat fortuitous because the starting point for the adjustment, i.e., the plane-wave LCAO-HF calculation, does not perfectly agree with the accurate B3LYP/TZVP molecular calculation. However, the need for and effect of the correction is clear.
As can be seen in table~\ref{table:ch4as}, a considerable narrowing is observed with the FWHM reduced from 2.52~keV to 2.09~keV, compared with the measured value of 2.06~keV.
Thus most of the required narrowing to obtain agreement with experiment has been accounted for through the effective inclusion of the nuclear repulsion with all of the atoms.
In the case of H$_2$ we saw that electron-positron annihilation vertex corrections further narrow he annihilation spectrum (see \fig{fig:g_factorH}). Therefore, it is expected that if one combined more accurate atomic correction factors with the broader LCAO-STO (or B3LYP/TZVP) electron momentum density (see \fig{fig:ch4as}), the resulting annihilation spectrum would provide a good description of the experimental data.

Finally, note that a broad core contribution can be seen clearly in the experimental spectrum for Doppler shifts $> 3$~keV. To reproduce this feature, the theory would need to be able to calculate accurately the size of the core contribution relative to that of the valence molecular orbitals~\cite{DGG_innershells}.

\begin{table}
\caption{\label{table:ch4as}FWHM of the $\gamma$-spectra for CH$_4$ (in keV).}
\begin{indented}
\item[]\begin{tabular}{cccccccc}
\br
Orbital		& B3LYP	& \multicolumn{3}{c}{LCAO} & Exp.\\
\cline{3-5}\noalign{\smallskip}
		&	 & STO & HF & HF$^{\rm a}$ & \\
\mr
$2a_1$				&1.96		& 1.85		&1.64		&1.45& --\\
$1t_2$				& 3.31		& 3.52		& 2.98		& 2.55& --\\
Total (valence) 			& 2.81		& 2.85		& 2.43	&2.09 & -- \\	
Total (all)		 & 2.97 	 	& 2.97	 	& 2.52	&2.09 &2.06\\	
\br
\end{tabular} 
\item[] $^{\rm a}$ Corrected by using the atomic annihilation momentum densities.
\end{indented}
\end{table}


\subsection{Gamma spectrum of tetrafluoromethane}\label{subsec:CF4}

Like methane, tetrafluoromethane (CF$_4$) belongs to the point group $T_d$. 
Similar to the hydrogens in CH$_4$, the groups of fluorine $1s$, $2s$ and $2p_z$ orbitals (with the $z$ axis directed locally towards the carbon atom) span the reducible representation $\Gamma^{\rm red}_{\sigma}=A_1\oplus T_2$. 
In contrast to CH$_4$, for CF$_4$ one has the additional complication of $\pi$ bonding, with the complete set of the eight fluorine $2p_{x,y}$ orbitals forming a basis for a reducible representation $\Gamma_{\pi}^{\rm red}=E \oplus T_1 \oplus T_2$\,(see e.g., \cite{cotton, DGG_thesis}). 
Accordingly, the B3LYP/TZVP calculation for CF$_4$ gives the following electronic orbital structure: core $+$ $1a_1^2$ $1t_2^6$ $2a_1^2$ $2t_2^6$ $e^4$ $3t_2^6$ $t_1^6$, where
the core consists of a triply degenerate $T_2$ symmetry orbital and a fully symmetric orbital that are predominantly F~$1s$ in nature, and another fully symmetric orbital which is predominantly C~$1s$ in nature.
Owing to the repulsion from the nuclei, the positron will have difficulty reaching and subsequently annihilating on these core orbitals, and the effect of the correction factors will be to greatly suppress the magnitude of the contribution of these core orbitals in comparison with the valence orbitals.
 
To employ the momentum-dependent correction factors, note that the $e$ and $t_1$ molecular orbitals are nonbonding, and are composed entirely of F~$2p$ orbitals. 
Their respective electron-momentum densities can therefore simply be multiplied by $G_{{\rm F}2p}(P)$ when calculating the corrected annihilation spectra. 
To apply the correction to the remaining molecular orbitals, however, we must construct the LCAO molecular orbitals and replace the electron momentum densities with the annihilation momentum densities calculated, in this case, in the Hartree-Fock zeroth-order vertex approximation, as was done for CH$_4$.
The molecular orbitals are 
\begin{eqnarray*}
\psi_{a_1} \!&=&\! \alpha_1 \psi_{C1s}+ \beta_1 \psi_{C2s}+ \gamma_1 {\sigma}_{{\rm F}1s}^{A_1} + \delta_1{\sigma}_{{\rm F}2s}^{A_1} +
\eta_1{\rm \sigma}_{{\rm F}2p_z}^{A_1}\\
\psi_{t_2x,y,z}\!&=&\! \alpha_2\psi_{{\rm C}2p} + \beta_2{\sigma}^{T_2}_{{\rm F}2sx,y,z}+ \gamma_2\sigma^{T_2}_{{\rm F}2p_{x,y,z}} +\delta_2\pi^{T_2}_{{\rm F}2px,y,z}
\end{eqnarray*}
where $\alpha $, $\beta$, $\gamma $, $\delta$ and $\eta$ are the expansion coefficients, and the $\sigma $ and $\pi $ SALC orbitals are given in Ref.~\cite{rozenbergcf4}.

\begin{figure}[t!]
\centering
\includegraphics*[width=0.8\textwidth]{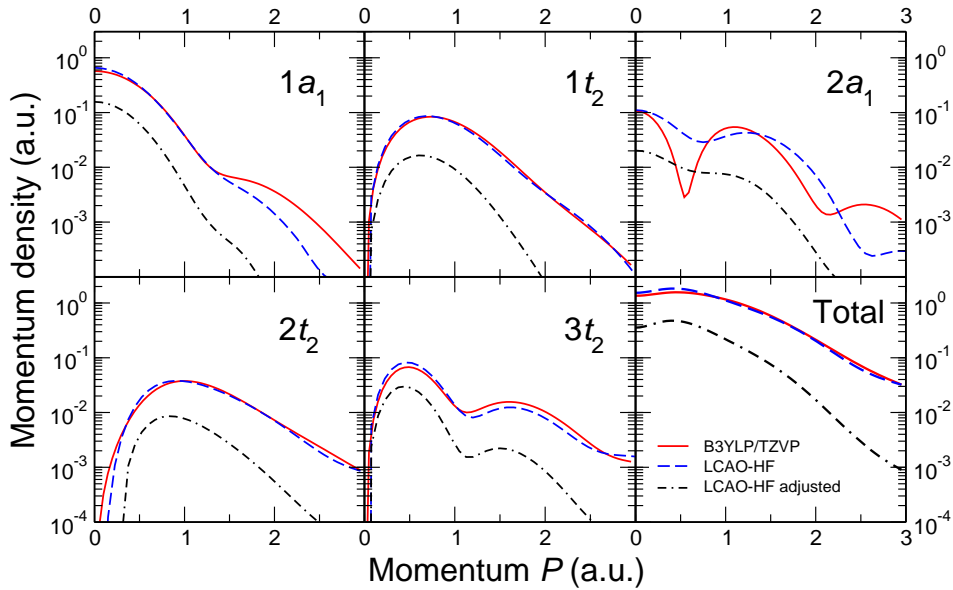}
\vspace{-3pt}
\caption{Electron momentum density of the valence orbitals of CF$_4$ calculated using the B3LYP (solid lines) and the LCAO-HF approach (dashed lines). Dot-dashed lines show the annihilation momentum densities
obtained in the LCAO-HF approach using atomic orbital correction factors.}
\label{fig:cf4dens}
\end{figure}

Figure \ref{fig:cf4dens} shows the momentum densities for the valence molecular orbitals calculated in the LCAO approach from \eqn{eqn:rhomolall} using the Hartree-Fock atomic orbitals. We determined the coefficients in the LCAO expansions by fitting the LCAO-HF momentum densities to the accurate B3LYP results (see \fig{fig:cf4dens}), using the coefficients from Ref.~\cite{rozenbergcf4} as the starting approximation. This approach yields the LCAO-HF electron momentum densities in good agreement with the B3LYP
results, especially for the $t_2$ orbitals which contain most of the electrons. The total valence electron momentum density obtained in the LCAO-HF approach is very close to that of the B3LYP calculation. Also shown in \fig{fig:cf4dens} are the annihilation momentum densities obtained in the LCAO-HF approach using atomic annihilation momentum densities from the zeroth-order Hartree-Fock calculation (i.e., including the $G_0^{\rm HF}(P)$ factors for the atomic orbitals involved). As expected, the correction factors reduce the magnitudes of the momentum densities (because of the positron repulsion from the nuclei) and, in particular, suppress the contribution of high momenta. 

\begin{figure}[t!]
\centering
\includegraphics*[width=0.6\textwidth]{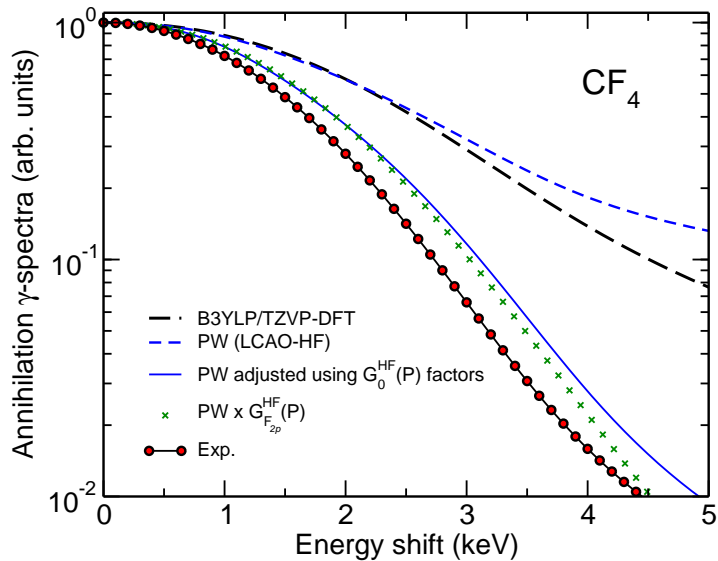}
\vspace{-3pt}
\caption[Annihilation $\gamma$-spectra of CF$_4$]{Annihilation $\gamma$-spectra of CF$_4$: (long-dashed line) B3LYP calculation; (short-dashed line) LCAO-HF (both correspond to the low-energy plane-wave positron approximation); (solid line) LCAO-HF with atomic correction factors; (crosses) LCAO-HF with the fluorine $2p$-orbital correction factor; (circles) experiment, two-Gaussian fit, Ref.~\cite{PhysRevA.55.3586}. The spectra were calculated from the total densities shown in \fig{fig:cf4dens} with power-type momentum density tails, and normalised to unity at $\epsilon =0$.}
\label{fig:cf4}
\end{figure}

Figure \ref{fig:cf4} shows the calculated annihilation $\gamma$-spectra
along with the the measured spectrum. The corresponding FWHM values are given in table~\ref{table:cf4}. The spectra from the B3LYP and LCAO-HF calculations, which neglect the positron-molecule interaction, are in good agreement with each other, but are significantly broader than the experiment. On the other hand, the LCAO-HF calculation that incorporates the atomic correction factors gives results much closer to the measured spectrum.
For CF$_4$, the dominant contribution to the annihilation momentum density comes from the twenty fluorine $2p$ electrons. 
Therefore, in addition, the figure shows the result of a simplified 
LCAO-HF calculation in which we use the correction factor of the fluorine $2p$ atomic orbital for \emph{all} molecular orbitals. 
The results of both adjustment methods are very similar. This suggest that if a specific atomic orbital dominates the molecular annihilation spectra, then the spectra can be brought in line with experiment using only the correction factor of this orbital.

\begin{table}[t!]
\caption{\label{table:cf4} FWHM of the $\gamma$-spectra of CF$_4$ (in keV).}
\begin{indented}
\item[]\begin{tabular}{ccccc}
\br
B3LYP		& \multicolumn{3}{c}{LCAO} 					& Exp.	\\
\cline{2-4}\noalign{\smallskip} 
			& HF 	& HF$^{\rm a}$ & HF$^{\rm b}$ 		& 		\\
\mr
4.48 			& 4.52 	& 3.32 		& 3.17 				& 2.92 	\\	
\br
\end{tabular}
\item[] $^{\rm a}$ Corrected by using the atomic annihilation momentum densities.
\item[] $^{\rm b}$ Corrected by using the fluorine $2p$-orbital factor $G_{{\rm F}2p}(P)$.
\end{indented}
\end{table}
%

The inclusion of the correction factors causes $\approx 30\%$ reduction in the linewidth, as documented in table~\ref{table:cf4}, and brings it into improved agreement with the measured value.
Note that one can gauge the relative importance of the inclusion of the atomic potential and the inclusion of the proper vertex in fluorine by considering these effects in neon. 
For the valence $2s$ and $2p$ orbitals in neon, the plane-wave positron linewidth of 4.94~keV~\cite{wangnobles}. This compares with the FWHM of 3.82~keV for the HF zeroth-order-vertex calculation, and 3.55~keV, when using the Dyson orbital and full 0th\,$+$\,1st-order\,$+\Gamma $ vertex \cite{0953-4075-39-7-008,Dunlop_thesis}. 
Thus for neon the majority of the narrowing comes from the inclusion the effect of the atomic potential, and we should also expect this to be the case for CF$_4$. The zeroth-order HF correction factor therefore brings
the calculation closer to the measured value than was observed for the corresponding quantity in H$_2$, where the nuclear potential is not as strong.


\section{Summary and conclusions}\label{sec:conclusions}

Two approaches have been used to calculate $\gamma$-spectra for positron annihilation for a selection of molecules including methane and its fluorosubstitutes.
The first method used modern quantum chemistry tools and density functional theory (B3LYP or BP86) to calculate the electron momentum densities, which were then converted into the annihilation spectra. This method assumes that a positron thermalized at room temperature can be described using the plane-wave approximation, i.e., neglecting its interaction with the target.
The corresponding annihilation spectra were found to be systematically broader than experiment by about 40\%.

These calculations were complemented in an approach that relied on expanding the molecular orbitals as linear combination of atomic orbitals. It allowed us to incorporate the effect of positron-atom interactions and electron-positron correlations using momentum-dependent atomic correction factors.
In particular, this method enables one to include the effect of positron-nuclear repulsion, which leads to a significant narrowing of the annihilaton spectra compared to the plane-wave approximation.

For the hydrogen atoms the atomic correction factor could be calculated near exactly from many-body theory. Its inclusion produced a $\gamma$-spectrum of H$_2$ in good agreement with experiment. Furthermore, for H$_2$ the method gave the absolute annihilation rate parameter $Z_{\rm eff}\approx 14$ for thermal room-temperature positrons. This value is close to the commonly accepted experimental $Z_{\rm eff}=14.6$ \cite{larrichia_h2} and best variational calculation, $Z_{\rm eff}=15.7$ \cite{PhysRevLett.103.223202}.

For carbon and fluorine we used much simpler, Hartree-Fock positron correction factors.
Nevertheless, the effective inclusion of the nuclear repulsion through these correction factors resulted in a significant narrowing of the spectra of CH$_4$ and CF$_4$, giving much improved agreement with experiment. For a polyatomic molecule the annihilation spectrum is determined by contributions of many molecular orbitals. In CF$_4$ the dominant orbitals are those that are fluorine-$2p$ in nature, which allows one to adjust the spectrum using a single atomic correction factor.
The remaining discrepancies are due to the neglect of the positron-atom and positron-electron correlations.

Any complete \emph{ab initio} theory of positron annihilation on molecules must include the full electron and positron dynamics, accounting for the positron-nuclear repulsion and electron-positron correlation effects. 
However, we have shown that realistic annihilation $\gamma$-spectra for molecules can be obtained from a standard LCAO electron momentum density approach, by replacing the atomic electron momentum densities (equivalent to plane-wave annihilation momentum densities) with atomic annihilation momentum densities.

We have shown that the effective distortion of the electron momentum density, when it is observed through positron annihilation $\gamma$-spectra, can be approximated by a simple scaling which reduces the momenta and $\gamma $-ray Doppler shifts by a factor of 1.4.
The origin of this scaling is in the similarity between the electron orbital momentum densities and corresponding momentum-dependent positron correction factors.
Realistic annihilation $\gamma$-spectra can therefore be obtained from standard electron momentum density calculations by including this scaling. 
Conversely, the simple scaling factor allows for the deduction of electron momentum densities from positron annihilation $\gamma$-spectra.

\ack
DGG was funded by a DEL Northern Ireland postgraduate studentship.
FW and SS acknowledge the support of the Australian Research Council under the Discovery Project scheme, and the National Computational Infrastructure at the Australian National University under the Merit Allocation Scheme. 
The research at the University of California, San Diego is supported by the NSF, Grant No.~PHY 10-68023.

\section*{References}

\begin{thebibliography}{10}

\bibitem{Dirac_annihilation}
Dirac P~A~M 1930 {\em Math. Proc. Cambridge Philos. Soc.} {\bf 26} 361

\bibitem{qed}
Berestetskii V~B, Lifshitz E~M and Pitaevskii L~P 1982 {\em Quantum
  Electrodynamics - 2nd ed. - (Course of Theoretical Physics; V4)} (Pergamon
  Press)

\bibitem{PhysRevA.55.3586}
Iwata K, Greaves R~G and Surko C~M 1997 {\em Phys. Rev. A} {\bf 55}
  3586

\bibitem{PhysRevLett.79.39}
Iwata K, Gribakin G~F, Greaves R~G and Surko C~M 1997 {\em Phys. Rev. Lett.}
  {\bf 79} 39

\bibitem{0953-4075-29-12-004}
Reeth P~V, Humberston J~W, Iwata K, Greaves R~G and Surko C~M 1996 {\em J.
  Phys. B} {\bf 29} L465

\bibitem{0953-4075-39-7-008}
Dunlop L~J~M and Gribakin G~F 2006 {\em J. Phys. B} {\bf 39} 1647

\bibitem{DGG_hlike}
Green D~G and Gribakin G~F 2012 Positron scattering and annihilation on
  hydrogen-like ions {T}o be submitted to Phys. Rev. A

\bibitem{DGG_innershells}
Green D~G and Gribakin G~F 2012 Enhancement and spectra of positron
  annihilation on core and valence electrons {T}o be submitted to Phys. Rev. A

\bibitem{wangnobles}
Wang F, Selvam L, Gribakin G~F and Surko C~M 2010 {\em J. Phys. B} {\bf 43}
  165207

\bibitem{RevModPhys.60.701}
Schultz P~J and Lynn K~G 1988 {\em Rev. Mod. Phys.} {\bf 60} 701

\bibitem{RevModPhys.66.841}
Puska M~J and Nieminen R~M 1994 {\em Rev. Mod. Phys.} {\bf 66} 841

\bibitem{PhysRevB.65.245310}
Saniz R, Barbiellini B and Denison A 2002 {\em Phys. Rev. B} {\bf 65} 245310

\bibitem{PhysRevB.66.041305}
Weber M~H, Lynn K~G, Barbiellini B, Sterne P~A and Denison A~B 2002 {\em Phys.
  Rev. B} {\bf 66} 041305

\bibitem{PhysRevB.68.165326}
Saniz R, Barbiellini B, Denison A~B and Bansil A 2003 {\em Phys. Rev. B} {\bf
  68} 165326

\bibitem{nmat1550}
Eijt S~W~H, van Veen A, Schut H, Mijnarends P~E, Denison A~B, Barbiellini B and
  Bansil A 2006 {\em Nat.~Mater.} {\bf 5} 23

\bibitem{eijt:091908}
Eijt S~W~H, Mijnarends P~E, van Schaarenburg L~C, Houtepen A~J, Vanmaekelbergh
  D, Barbiellini B and Bansil A 2009 {\em Appl. Phys. Lett.} {\bf 94} 091908


\bibitem{meng:093510}
Meng X~Q, Chen Z~Q, Jin P, Wang Z~G and Wei L 2007 {\em Appl. Phys. Lett.}
  {\bf 91} 093510

\bibitem{PhysRevB.79.201405}
Nagai Y, Toyama T, Tang Z, Inoue K, Chiba T, Hasegawa M, Hirosawa S and Sato T
  2009 {\em Phys. Rev. B} {\bf 79} 201405

\bibitem{PhysRevB.73.014111}
Maekawa M, Kawasuso A, Yoshikawa M, Miyashita A, Suzuki R and Ohdaira T 2006
  {\em Phys. Rev. B} {\bf 73} 014111

\bibitem{positronnanocrystals}
Eijt S~W~H, Barbiellini B, Houtepen A~J, Vanmaekelbergh D, Mijnarends P~E and
  Bansil A 2007 {\em Phys. Status Solidi ({C})} {\bf 4} 3883

\bibitem{PhysRevLett.77.2097}
Asoka-Kumar P, Alatalo M, Ghosh V~J, Kruseman A~C, Nielsen B and Lynn K~G 1996
  {\em Phys. Rev. Lett.} {\bf 77} 2097

\bibitem{PhysRevLett.82.3819}
Petkov M~P, Weber M~H, Lynn K~G, Crandal~l R~S and Ghosh V~J 1999 {\em Phys.
  Rev. Lett.} {\bf 82} 3819

\bibitem{PhysRevB.51.4176}
Alatalo M, Kauppinen H, Saarinen K, Puska M~J, M\"akinen J, Hautoj\"arvi P and
  Nieminen R~M 1995 {\em Phys. Rev. B} {\bf 51} 4176

\bibitem{PhysRevB.73.014114}
Kim S, Eshed A, Goktepeli S, Sterne P~A, Koymen A~R, Chen W~C and Weiss A~H
  2006 {\em Phys. Rev. B} {\bf 73} 014114

\bibitem{paescopper}
Mayer J, Schreckenbach K and Hugenschmidt C 2007 {\em Phys. Status Solidi
  ({C})} {\bf 4} 3928

\bibitem{RevModPhys.82.2557}
Gribakin G~F, Young J~A and Surko C~M 2010 {\em Rev. Mod. Phys.} {\bf 82}
  2557

\bibitem{PhysRevLett.57.1651}
Brown B~L and Leventhal M 1986 {\em Phys. Rev. Lett.} {\bf 57} 1651

\bibitem{PhysRevLett.68.3793}
Tang S, Tinkle M~D, Greaves R~G and Surko C~M 1992 {\em Phys. Rev. Lett.}
  {\bf 68} 3793

\bibitem{0953-4075-30-6-025}
Armour E~A~G and Carr J~M 1997 {\em J. Phys. B} {\bf 30} 1611

\bibitem{darewych}
Darewych J~W 1979 {\em Can. J. Phys.} {\bf 57} 1027

\bibitem{ghosh}
Ghosh A~S, Mukherjee T and Darewych J~W 1994 {\em Hyperfine Interact.} {\bf
  89} 319

\bibitem{chuang}
Chuang S~Y and Hogg B~G 1967 {\em Can. J. Phys.} {\bf 45} 3895

\bibitem{fluorobenzenes}
Wang F, Ma X, Selvam L, Gribakin G and Surko C~M 2012 Chemical structural
  effects on $\gamma $-ray spectra of positron annihilation in fluorobenzenes
  {S}ubmitted to Eur. Phys. J. D
  
\bibitem{GSW11}Green D~G, Saha S, Wang F, Gribakin G~F and Surko C~M 2011 {\em Material Science Forum} {\bf 666} 21

\bibitem{Akhiezer}
Akhiezer A~I and Berestetskii V~B 1982 {\em Quantum Electrodynamics}
  (Interscience Publishers 1965)

\bibitem{PhysRevA.70.032720}
Gribakin G~F and Ludlow J 2004 {\em Phys. Rev. A} {\bf 70} 032720

\bibitem{0953-4075-35-2-311}
Gribakin G~F and Ludlow J 2002 {\em J. Phys. B} {\bf 35} 339

\bibitem{emsreview}
McCarthy I~E and Weigold E 1991 {\em Rep. Prog. Phys.} {\bf 54} 789

\bibitem{Fraser}
Fraser P~A 1968 {\em Adv. At. Mol. Phys.} {\bf 4} 63

\bibitem{pomeranchuk}
Pomeranchuk I 1949 {\em Zh. Eksp. Teor. Fiz.} {\bf 19} 183

\bibitem{becke}
Becke A~D 1993 {\em J. Chem. Phys.} {\bf 98} 5648

\bibitem{PhysRevB.37.785}
Lee C, Yang W and Parr R~G 1988 {\em Phys. Rev. B} {\bf 37} 785

\bibitem{TZVP}
Godbout N and Salhub D~R 1992 {\em Can. J. Chemistry} {\bf 70} 560

\bibitem{fengemds}
Wang F 2003 {\em J. Phys. Chem. A} {\bf 107} 10199

\bibitem{g03}
Gaussian 03 M~J~Frisch et~al{$,$} 2004 \uppercase{G}aussian, Inc., Wallingford, CT

\bibitem{GAMESS}
Schmidt M~W, Baldridge K~K, Boatz J~A, Elbert S~T, Gordon M~S, Jensen J~H,
  Koseki S, Matsunaga N, Nguyen K~A, Su S, Windus T~L, Dupuis M and Montgomery
  J~A 1993 {\em J. Comput. Chem.} {\bf 14} 1347

\bibitem{Kaijser197737}
Kaijser P and Jr V~H~S 1977 {\em Adv. Quantum Chem.} {\bf 10} 37

\bibitem{varshalovich}
Varshalovich D~A, Moskalev A~N and Khersonskii V~K 1988 {\em Quantum theory of
  angular momentum} (World Scientific)

\bibitem{slater}
Slater J~C 1963 {\em Quantum Theory of Molecules and Solids} (McGraw-Hill
  Book Company, Inc.)

\bibitem{DGG_thesis}
Green D~G 2011 {\em Positron annihilation with core electrons} Ph.D. thesis
  Queen's University Belfast

\bibitem{armourh2_0}
Armour E~A~G and Baker D~J 1986 {\em J. Phys. B} {\bf 19} L871

\bibitem{armourh2_1}
Cooper J~N, Armour E~A~G and Plummer M 2008 {\em J. Phys. B} {\bf 41} 245201

\bibitem{armourh2_new}
Armour E~A~G, Cooper J~N, Gregory M~R, Jonsell S, Plummer M and Todd A~C 2010
  {\em J. Phys. Conf. Ser.} {\bf 199} 012007

\bibitem{PhysRevLett.103.223202}
Zhang J~Y, Mitroy J and Varga K 2009 {\em Phys. Rev. Lett.} {\bf 103}
  223202

\bibitem{ruizhangh2}
Zhang R, Baluja K~L, Franz J and Tennyson J 2011 {\em J. Phys. B} {\bf 44}
  035203

\bibitem{PhysRevA.20.347}
McNutt J~D, Sharma S~C and Brisbon R~D 1979 {\em Phys. Rev. A} {\bf 20}
  347

\bibitem{charlton_h2}
Wright G~L, Charlton M, Clark G, Griffith T~C and Heyland G~R 1983 {\em J.
  Phys. B} {\bf 16} 4065

\bibitem{larrichia_h2}
Laricchia G, Charlton M, Beling C~D and Griffith T~C 1987 {\em J. Phys. B}
  {\bf 20} 1865

\bibitem{cotton}
Cotton F~A 1990 {\em Chemical applications of Group Theory} 3rd ed (Wiley
  Interscience)

\bibitem{pitzer}
Pitzer R~M 1967 {\em J. Chem. Phys.} {\bf 46} 4871

\bibitem{rozenbergcf4}
Rozenberg E~L and Dyatkina M~E 1971 {\em J. Struct. Chem.} {\bf 12} 270

\bibitem{Dunlop_thesis}
Dunlop L~J~M 2005 {\em Theory of positron-atom annihilation gamma spectra}
  Ph.D. thesis Queen's University Belfast

\end{thebibliography}

\end{document}